\documentclass[11pt,a4paper]{article}
\usepackage[margin=24mm]{geometry}
\usepackage[T1]{fontenc}
\usepackage{lmodern,microtype,amsmath,amssymb,mathtools,bm,mathrsfs}
\usepackage{graphicx,booktabs,longtable,array,enumitem}
\usepackage{tikz}
\usetikzlibrary{arrows.meta,positioning,shapes.geometric}
\usepackage[numbers,sort&compress]{natbib}
\usepackage[colorlinks=true,linkcolor=blue,citecolor=blue,urlcolor=blue]{hyperref}
\usepackage[section]{placeins}
\usepackage{fancyhdr}
\allowdisplaybreaks[2]
\newcommand{\dd}{\mathrm d}
\newcommand{\bg}{\bar g}
\newcommand{\bn}{\bar\nabla}
\newcommand{\bR}{\bar R}
\newcommand{\bBox}{\bar\Box}
\newcommand{\GI}{\mathrm{GI}}
\newcommand{\Lie}{\mathcal L}
\newcommand{\Ric}{\mathcal R}

\newcommand{\Id}{\mathbf 1}

\newcommand{\ahat}{\widehat\alpha}
\newcommand{\AuthorName}{Soham Bhattacharyya}
\title{Static tidal Love numbers of Schwarzschild and\\non-Schwarzschild black holes in quadratic gravity}
\author{
  \AuthorName
  \thanks{Email:
    \href{xeonese@gmail.com}
         {\texttt{xeonese@gmail.com}}
  }\\[0.4em]
  \small Department of Physics, School of Sciences\\
  \small National Institute of Technology Andhra Pradesh\\
  \small Tadepalligudem 534101, India
}
\date{}

\hypersetup{pdftitle={Static quadrupolar tidal Love numbers of black holes in quadratic gravity}}
\begin{document}
\maketitle
\begin{abstract}
We calculate the static electric quadrupolar response of black holes in
four-dimensional quadratic gravity, with action $R-\alpha C^2+\beta R^2$.
The Schwarzschild solution has zero massless gravitational Love number
under the stated boundary conditions. A numerical non-Schwarzschild
solution instead has a nonzero negative response. At $\alpha/r_h^2=1/2$,
we find $\lambda_E/r_h^5\simeq-0.1224928$ and
$k_2^{(r_h)}\simeq-0.1837391$, with a final outer-endpoint change in
$k_2^{(r_h)}$ of about $1.60\times10^{-7}$. We explain how the background
is obtained by shooting and how the tidal solution is selected by
matching at an intermediate radius. The perturbations are written in
Kodama--Ishibashi gauge-invariant variables. After the scalar-curvature
perturbation is excluded by horizon regularity and the absence of an
applied scalar tide, the remaining static sector has three coupled
second-order radial equations. We define the curvature amplitudes in
these equations and describe why simple attempts to separate them fail.
The coupling scan shows a negative response approaching zero near the
Schwarzschild intersection. We also explain why this full-theory result
does not conflict with the removal of curvature-squared operators in a
perturbative low-energy expansion. 
\end{abstract}
\tableofcontents
\clearpage

\section{Introduction}\label{sec:intro}
A black hole in an external gravitational field need not remain spherical.
The external field changes the geometry, and the change may contain an
induced mass quadrupole. The tidal Love number measures this response.
For a four-dimensional Schwarzschild black hole in general relativity,
the static gravitational Love number is zero
\cite{Binnington:2009bb,Damour:2009vw}. This result has been obtained
using metric perturbations, symmetry arguments and scattering amplitudes
\cite{Kol:2011vg,Chia:2020yla,Charalambous:2021kcz,Hui:2021vcv,Ivanov:2022qqt}.
Related calculations also find a vanishing response at higher orders
in the applied tidal field \cite{Riva:2023rcm,Iteanu:2024dvx}.
It is therefore natural to ask which parts of this result survive when
the gravitational field equations are changed.

We study this question in four-dimensional quadratic gravity,
\begin{equation}\label{eq:action}
 S=\frac{1}{16\pi G}\int\dd^4x\sqrt{-g}
 \left[R-\alpha C_{\mu\nu\rho\sigma}C^{\mu\nu\rho\sigma}
                  +\beta R^2\right].
\end{equation}
Here $C_{\mu\nu\rho\sigma}$ is the Weyl tensor. The constants
$\alpha$ and $\beta$ have dimensions of length squared. In addition
to the usual massless graviton, the theory contains a massive spin-two
field and a massive scalar \cite{Stelle:1976gc,Stelle:1977ry,Salvio:2018crh}.
With our sign convention, their masses are
\begin{equation}\label{eq:masses}
 m_2^2=\frac{1}{2\alpha},\qquad m_0^2=\frac{1}{6\beta}.
\end{equation}
The massive spin-two field has a ghost kinetic term. We retain this
field in the classical equations. We do not assume that doing so gives
a unitary quantum theory. Proposed quantum interpretations and the
classical calculations discussed here address different questions
\cite{Salvio:2018crh,Holdom:2016xfn}.

The Schwarzschild metric solves the full equations for any $\alpha$
and $\beta$. However, the theory also admits static, spherical black
holes that are not Schwarzschild. The numerical solutions of L\"u,
Perkins, Pope and Stelle provide an important example
\cite{Lu:2015cqa,Lu:2015psa}. Their Ricci scalar vanishes,
$\bar R=0$, but their Ricci tensor does not,
$\bar R_{\mu\nu}\ne0$. The distinction matters: a vanishing Ricci
scalar does not make these geometries vacuum Einstein solutions.

\subsection{What is known about black holes in quadratic gravity}
Work on quadratic gravity has addressed the existence of its solutions,
their stability and their gravitational-wave signals. Early studies
examined the Schwarzschild solution in fourth-order gravity
\cite{Whitt:1985ki,Nelson:2010ig}. Later work found the non-Schwarzschild branch,
related its intersection with Schwarzschild to a Lichnerowicz mode,
and developed other ways to describe spherical solutions
\cite{Lu:2015cqa,Lu:2015psa,Lu:2017kzi,Bonanno:2019rsq,Podolsky:2019gro,Svarc:2018coe}.
Analytical approximations to the numerical metric have also been
constructed \cite{Kokkotas:2017zwt}. A recent preprint reports a
computer-assisted existence proof for a non-Schwarzschild solution
near the case $m_2r_h=1$ \cite{Goldstein:2026svk}. These developments
provide several ways to check the background geometry.

Stability requires a separate perturbation calculation. The massive
spin-two sector of Schwarzschild has a long-wavelength instability
related to the Gregory--Laflamme instability of a black string
\cite{Brito:2013wya,Gregory:1993vy}. The Lichnerowicz negative mode
helps explain the relation between the instability threshold and the
intersection of the two black-hole branches
\cite{Gross:1982cv,Lu:2017kzi}. Held and Zhang found a corresponding radial
instability on the non-Schwarzschild branch \cite{Held:2022abx}.
Other analyses study different limits or theories; for example, a
Weyl-dominated coupling limit and the scale-invariant $R^2$ model
\cite{DeFelice:2023kpl,Dioguardi:2020nxr}. Their assumptions should not be
identified with those of the finite-coupling black holes used here.

Quasinormal-mode calculations ask how a disturbed black hole relaxes
in time. Early studies examined test fields on the non-Schwarzschild
background \cite{Cai:2015fia,Zinhailo:2018ska,Konoplya:2022iyn}. A test field probes the
geometry, but it does not include all the gravitational degrees of
freedom. Antoniou, Gualtieri and Pani calculated axial gravitational
quasinormal modes of both black-hole branches
\cite{Antoniou:2024jku}. Konoplya, Spina and Zhidenko studied their
time evolution, including the late-time massive tails
\cite{Konoplya:2025afm}. Recent work has also treated perturbations
of Kerr and radiation produced by particles falling into Schwarzschild
\cite{Knoska:2025jdj,Antoniou:2026nhh}. In particular, the latter study
shows how extra radiation can be exponentially suppressed as the
Einstein limit is approached \cite{Antoniou:2026nhh}.

The classical initial-value problem and nonlinear evolution have also
received attention \cite{Noakes:1983xd,Held:2021pht,Held:2023aap}.
Nonlinear calculations study the backreaction of unstable massive
fields and binary black-hole waveforms
\cite{East:2023nsk,Held:2025ckb}. These results show why the full
quadratic theory should not be confused with a calculation that keeps
only small corrections to Einstein gravity. They also motivate a
complementary question: what is the static response to a specified
external gravitational tide? A static response coefficient, an
instability growth rate and a quasinormal frequency are different
properties of the same perturbation problem
\cite{Cardoso:2017cfl,Barbosa:2025uau}.

\subsection{Why field redefinitions seem to give a different answer}\label{sec:motivation}
There is a simple reason to expect no leading curvature-squared
correction in a vacuum low-energy calculation. In four dimensions,
\begin{equation}
 C^2=\mathcal G+2R_{\mu\nu}R^{\mu\nu}-\frac23R^2,
 \qquad
 \mathcal G=R_{\mu\nu\rho\sigma}R^{\mu\nu\rho\sigma}
              -4R_{\mu\nu}R^{\mu\nu}+R^2.
\end{equation}
Thus the bulk Lagrangian is
\begin{equation}\label{eq:redundant}
 R+aR_{\mu\nu}R^{\mu\nu}+bR^2-\alpha\mathcal G,
 \qquad a=-2\alpha,\quad b=\beta+\frac{2\alpha}{3}.
\end{equation}
Multiply the quadratic terms by a formal expansion parameter
$\epsilon$. Consider the change of metric
\begin{align}\label{eq:redefinition}
 g_{\mu\nu}&=\widetilde g_{\mu\nu}
              +\epsilon\Delta g_{\mu\nu}+O(\epsilon^2),\\
 \Delta g_{\mu\nu}
 &=a\widetilde R_{\mu\nu}
       -\left(b+\frac a2\right)\widetilde g_{\mu\nu}\widetilde R
 =-2\alpha\widetilde R_{\mu\nu}
       +\left(\frac\alpha3-\beta\right)
                              \widetilde g_{\mu\nu}\widetilde R.
\end{align}
The change in the Einstein--Hilbert action is
\begin{align}\label{eq:EHvariation}
 S_{\rm EH}[\widetilde g+\epsilon\Delta g]
 &=S_{\rm EH}[\widetilde g]
 -\frac{\epsilon}{16\pi G}\int\dd^4x\sqrt{-\widetilde g}\,
                  \widetilde G^{\mu\nu}\Delta g_{\mu\nu}
 +O(\epsilon^2),\\
 -\widetilde G^{\mu\nu}\Delta g_{\mu\nu}
 &=-a\widetilde R_{\mu\nu}\widetilde R^{\mu\nu}
                                      -b\widetilde R^2.\nonumber
\end{align}
The two Ricci terms therefore cancel at this order. The remaining
Euler term does not change the local vacuum equations in four
dimensions, once the boundary terms are treated consistently.
This is a perturbative statement about redundant operators
\cite{Kamefuchi:1961sb,Deser:1986xr,deRham:2020ejn,Bhattacharyya:2022dzh}.
It applies to the Einstein branch of the low-energy expansion.

Three points are needed to interpret this result. First, the
transformed action has terms of order $\epsilon^2$ and higher.
Discarding them is an approximation. The transformation does not
turn the full finite-coupling theory into pure Einstein gravity.
Second, matter couplings change with the metric. For a point particle,
\begin{equation}\label{eq:worldlineshift}
 S_{\rm pp}=-m\int\dd\tau,
 \qquad
 \delta S_{\rm pp}=\frac{m\epsilon}{2}
       \int\dd\widetilde\tau\,
       \widetilde u^\mu\widetilde u^\nu\Delta g_{\mu\nu}.
\end{equation}
An observer in the redefined variables must use the transformed
couplings to infer the same physical geometry. Third, a nonlinear
black-hole solution may have no regular expansion in the coupling
around the chosen Einstein solution. The non-Schwarzschild branch
has structure on the length scale $m_2^{-1}=\sqrt{2\alpha}$.
Its Yukawa dependence, $e^{-r/\sqrt{2\alpha}}$, is not a power series
in $\alpha$ at fixed $r$.

An exact second-order formulation introduces the extra scalar and
tensor fields explicitly \cite{Hindawi:1995an}. It keeps their equations
and is therefore different from dropping the extra solutions by
low-energy order reduction. Likewise, a worldline effective field
theory replaces a compact object by localized operators; it need
not remove the extra fields from the bulk theory
\cite{Goldberger:2004jt,Porto:2016pyg,Kulkarni:2024ghc}.
We use this distinction throughout the paper.

There is consequently no disagreement between equivalent field
variables when the same solution, sources and measured quantities
are compared. The zero/nonzero question here concerns two black-hole
branches and two uses of the quadratic action. We calculate a metric
response in the original, minimally coupled metric. We do not identify
it with an independently matched worldline Wilson coefficient.
Higher-curvature matching can require extra localized contributions
\cite{Wang:2026qst}; that calculation is left for future work.
Appendix~\ref{app:frames} gives further details.

\subsection{Why we use a numerical black-hole background}
Bhattacharyya et al.\ \cite{Bhattacharyya:2025slf} reconstruct
asymptotic metrics and study tidal tails in quadratic gravity using
worldline effective field theory. Their results help motivate the
present question, but an asymptotic metric is not enough to impose
a regular black-hole boundary condition.

A weak-field expansion is controlled where $GM/r\ll1$. A horizon
lies in the strong-field region. Setting a truncated lapse to zero
does not establish a regular horizon of the full theory. Moreover,
the amplitudes of the asymptotic Yukawa terms must be those selected
by a regular interior boundary. They cannot be chosen independently
when defining a particular black-hole solution. For positive scalar
mass squared, a regular static asymptotically flat vacuum black hole
also obeys $R=0$ \cite{Lu:2015cqa,Lu:2015psa}. A scalar charge
allowed in a far-field reconstruction therefore needs a separate
regularity check before it is interpreted as black-hole hair.

These points do not prove that every asymptotic metric in
Ref.~\cite{Bhattacharyya:2025slf} lacks a black-hole completion.
They explain why we use a global numerical solution, with a verified
horizon and asymptotic region, for the present tidal calculation.

\subsection{Plan of the paper}
We consider static, electric, $\ell=2$ perturbations that are even
under time reflection. The applied tide is massless and gravitational.
There is no applied scalar or growing massive tensor tide. Decaying
massive responses are allowed. We take $\alpha>0$ and $\beta\geq0$,
and use $G=c=1$ in numerical results.

Section~\ref{sec:background} explains how the numerical background
is constructed. Section~\ref{sec:perturbations} defines the
Kodama--Ishibashi variables and derives the three coupled radial
equations. It also gives the zero response on Schwarzschild.
Section~\ref{sec:TLN} explains the matching calculation and presents
the Love numbers and coupling scan. We discuss their meaning and
the remaining questions in Sec.~\ref{sec:conclusions}. The
appendices contain the full coefficient arrays and longer derivations.

\section{Constructing the black-hole background}\label{sec:background}
\subsection{The geometry and its parameters}
Write the spherical metric in areal radius as
\begin{equation}\label{eq:metric}
 \dd\bar s^2=-A(r)\dd t^2+\frac{\dd r^2}{B(r)}+r^2\dd\Omega^2.
\end{equation}
The horizon is a common simple zero of $A$ and $B$. Both functions
are positive outside it, and we normalize $A(\infty)=B(\infty)=1$.
The full trace equation is
\begin{equation}\label{eq:trace}
 6\beta\Box R-R=0.
\end{equation}
For a regular static asymptotically flat solution with $\beta>0$,
an integration argument gives $\bar R=0$. For $\beta=0$ this follows
directly from the trace. The remaining background equations are
\begin{equation}\label{eq:bg}
 (\bBox-m_2^2)\bR_{\mu\nu}
 +2\bR_{\mu\rho\nu\sigma}\bR^{\rho\sigma}
 -\frac12\bg_{\mu\nu}\bR_{\rho\sigma}\bR^{\rho\sigma}=0,
 \qquad\bR=0.
\end{equation}
They contain no $\beta$. Schwarzschild has $\bar R_{\mu\nu}=0$;
the non-Schwarzschild solution has $\bar R_{\mu\nu}\ne0$.

We label the solutions by
\begin{equation}\label{eq:parameters}
 p=m_2r_h,\qquad\ahat=\frac{\alpha}{r_h^2}=\frac{1}{2p^2},
 \qquad B'(r_h)=\frac{1+\delta}{r_h}.
\end{equation}
The number $\delta$ measures the change in the radial horizon slope
from Schwarzschild. It is the shooting parameter. The choice
$\delta=0$ gives Schwarzschild; a different root of the asymptotic
condition selects the non-Schwarzschild branch.

\subsection{How the shooting method works}\label{sec:shooting}
The boundary conditions are imposed at two ends of the exterior:
regularity at the horizon and flatness at infinity. A direct
integration cannot start exactly at the horizon because the radial
equations contain coefficients such as $1/B$. We first construct
a regular power series there.

Let $\xi=(r-r_h)/r_h$. Before fixing the time normalization, write
\begin{equation}\label{eq:bghorizon}
 A=a_1\xi+a_2\xi^2+\cdots,\qquad
 B=(1+\delta)\xi+b_2\xi^2+\cdots.
\end{equation}
The freedom to rescale $t$ allows us to set $a_1=1$ at this stage.
For chosen $p$ and $\delta$, substitution into the field equations
determines the higher coefficients. Evaluating the series at a small
positive $\xi=\epsilon_b$ gives nonsingular initial data for the
outward integration. The value of $\delta$ is still to be found.

At large radius the extra tensor field permits both growing and
decaying Yukawa terms. A generic trial $\delta$ excites a growing
term. It eventually spoils flatness, even when its initial amplitude
is very small. Shooting means changing $\delta$ until the growing
term is removed. This is why the horizon slope cannot be chosen
freely for an asymptotically flat black hole.

We use a finite outer radius $R$ to measure this mismatch. In the
dimensionless coordinate $z=m_2r$, define
$\widehat\rho_t=\bar R^t{}_t/m_2^2$ and
$\mathcal M_Z=Z[1-B(Z)]/2$, where $Z=m_2R$. The leading decaying
tail is $\widehat\rho_t\sim e^{-z}z^{-1-m_2M}$. We therefore use
the shooting residual
\begin{equation}\label{eq:shootresidual}
 \mathcal F(\delta;Z)=
 \left.\frac{\dd\widehat\rho_t}{\dd z}\right|_Z
 +\left(1+\frac{1+\mathcal M_Z}{Z}\right)
                                \widehat\rho_t(Z).
\end{equation}
This residual is approximately zero for decaying hair at a large
finite radius. It is not an exact condition at infinity.

\begin{figure}[tbp]
 \centering\begin{tikzpicture}[
  font=\small, node distance=7mm,
  box/.style={draw=black!65,rounded corners=2pt,fill=blue!3,
              align=center,text width=7.4cm,minimum height=9mm,inner sep=4pt},
  choice/.style={diamond,draw=black!65,aspect=3.3,align=center,
                 text width=4.3cm,inner sep=1pt,fill=orange!7},
  flow/.style={-{Stealth[length=2mm]},thick,draw=black!70}]
 \node[box] (choose) {Choose $p=m_2r_h$ and a trial horizon slope $\delta$};
 \node[box,below=of choose] (series)
 {Build the regular horizon series\\and start at $r_h+\epsilon_b$};
 \node[box,below=of series] (integrate)
 {Integrate the nonlinear equations to $R$};
 \node[choice,below=of integrate] (residual)
 {Is the growing-tail residual\\$\mathcal F(\delta;m_2R)$ small enough?};
 \node[box,text width=2.9cm,right=10mm of residual] (adjust)
 {Change $\delta$\\while following the\\non-Schwarzschild root};
 \node[box,below=of residual] (normalise)
 {Normalize $A$ at infinity and estimate $M_R$};
 \node[box,below=of normalise] (check)
 {Repeat with larger $R$ and smaller $\epsilon_b$;\\check the exterior and the original field equations};
 \draw[flow] (choose)--(series);
 \draw[flow] (series)--(integrate);
 \draw[flow] (integrate)--(residual);
 \draw[flow] (residual)--node[above,font=\footnotesize]{no}(adjust);
 \draw[flow] (adjust.north)|-(series.east);
 \draw[flow] (residual)--node[right,font=\footnotesize]{yes}(normalise);
 \draw[flow] (normalise)--(check);
\end{tikzpicture}
 \caption{Shooting for an asymptotically flat non-Schwarzschild
 background. The adjustable quantity is the horizon slope $\delta$.
 The residual tests for the unwanted growing massive term at a finite
 outer radius. Repeating the calculation with larger $R$ and smaller
 horizon offset checks the finite-boundary approximation.}
 \label{fig:shooting}
\end{figure}
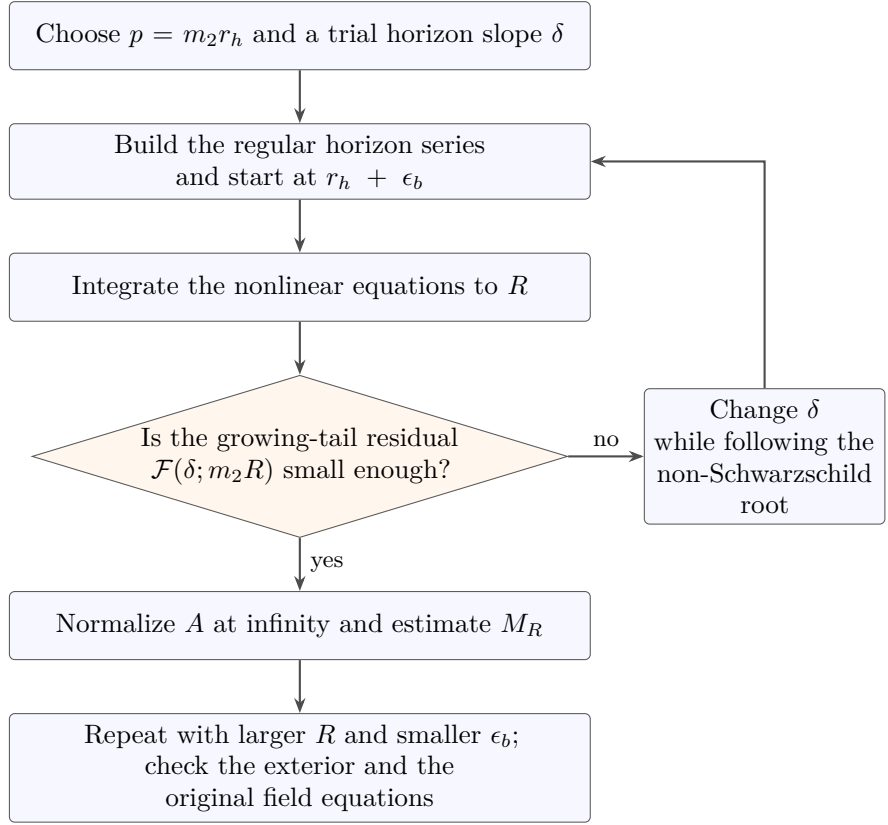

For each trial $\delta$, we integrate the nonlinear equations and
evaluate $\mathcal F$. We locate the nonzero root by varying the
trial slope. We keep track of the branch as $p$ changes; otherwise
the root finder can return the known Schwarzschild solution.
A bracket between trial values with opposite residual signs is useful
when such a bracket has been found. The residual need not be monotonic
over an arbitrary range, so identifying the branch is part of the
calculation. Figure~\ref{fig:shooting} shows the procedure.

After a root has been found, we divide the raw $A$ by its estimated
asymptotic normalization. We estimate the mass from
\begin{equation}\label{eq:massestimate}
 M_R=\frac R2[1-B(R)].
\end{equation}
We then increase $R$, reduce the horizon offset, and change the series
order or working precision. A reliable background must retain a
regular exterior, a stable mass estimate, and small residuals in the
original field equations. A very accurate root of the finite-radius
condition alone is not sufficient.

For the integration we propagate the mixed Ricci components together
with the metric functions. This avoids a removable denominator
$rA'-2A$ that appears when one solves directly for $B''$. The full
equations and series prescription are in Appendix~\ref{app:background}.
This numerical construction follows the general horizon-to-infinity
strategy used in studies of the non-Schwarzschild branch
\cite{Lu:2015cqa,Lu:2015psa,Bonanno:2019rsq}.

\subsection{The background used for the reference calculation}
Our reference case is $p=1$, or $\ahat=1/2$. Set $r_h=m_2=1$.
The recorded refined solution has
\begin{equation}\label{eq:baselinebg}
 \delta\simeq0.3633018786,\qquad
 M/r_h\simeq0.2755350525.
\end{equation}
The quoted mass is the estimator at $R/r_h=40$.
Before time normalization, the asymptotic lapse estimate is
$0.6216822438$. The horizon invariant is
\begin{equation}
 r_h^4(\bar R_{\mu\nu}\bar R^{\mu\nu})_h\simeq0.5279530.
\end{equation}
It is nonzero even though $\bar R=0$.

\begin{figure}[tbp]
 \centering
 \includegraphics[width=.96\linewidth]{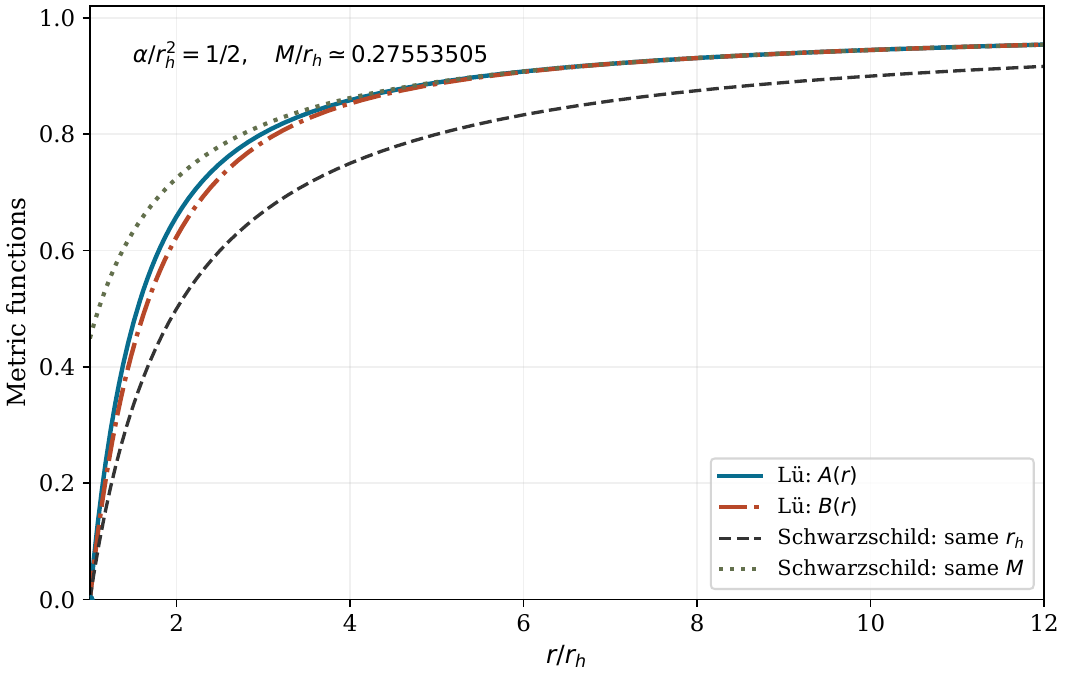}
 \caption{The metric functions at $\alpha/r_h^2=1/2$. The two
 Schwarzschild curves have either the same horizon radius or the same
 asymptotic mass as the non-Schwarzschild solution. Both comparisons
 are shown because $r_h\ne2M$ on this branch. The plot uses numerical
 background samples, with $A$ normalized at infinity.}
 \label{fig:bg}
\end{figure}

The equal-horizon Schwarzschild solution has $M/r_h=1/2$, whereas
the reference non-Schwarzschild black hole has a smaller mass.
The equal-mass Schwarzschild solution instead has its horizon at
$2M<r_h$. Figure~\ref{fig:bg} makes these two comparisons separately
on the same axes. They must not be confused when describing how
close a numerical metric is to Schwarzschild.

The familiar positive-mass part of the L\"u family extends
approximately from $p\simeq0.876$ at the Schwarzschild intersection
to $p\simeq1.143$ at zero mass
\cite{Lu:2015cqa,Lu:2017kzi,Held:2022abx}.
At fixed $r_h$ these correspond to $\ahat\simeq0.65$ and
$\ahat\simeq0.383$. They describe this family, not bounds on all
solutions of quadratic gravity. A change in the sign of $\delta$
does not, by itself, establish a singularity or exclude another
continuation. The global equations and boundary conditions must
be checked again.

\section{From gauge invariants to three coupled equations}\label{sec:perturbations}
\subsection{Angular harmonics and the seven metric amplitudes}
We split the background into a two-dimensional orbit space and a unit
sphere. Lower-case Latin indices $a,b$ refer to $(t,r)$; upper-case
Latin indices $A,B$ refer to the angular coordinates. The orbit metric
is $g_{ab}\dd x^a\dd x^b=-A\dd t^2+\dd r^2/B$.
Let $D_a$ and $D_A$ be the covariant derivatives on these two spaces.
For a scalar spherical harmonic $Y=Y_{\ell m}$, define
\begin{equation}\label{eq:harmonics}
 D^AD_AY=-k^2Y,\qquad k^2=\ell(\ell+1),\qquad
 Y_A=D_AY,\qquad
 Y_{AB}=D_AD_BY+\frac{k^2}{2}\gamma_{AB}Y.
\end{equation}
The tensor $Y_{AB}$ is trace free. We set $\ell=2$ from now on,
so $k^2=6$ and $Y_{AB}=D_AD_BY+3\gamma_{AB}Y$.

The stationary polar metric perturbation has seven radial amplitudes:
\begin{align}\label{eq:seven}
 h_{tt}&=AH_0Y,& h_{tr}&=H_1Y,&h_{rr}&=H_2Y/B,\nonumber\\
 h_{tA}&=j_0Y_A,&h_{rA}&=JY_A,&
 h_{AB}&=r^2(K\gamma_{AB}Y+2G_{\rm met}Y_{AB}).
\end{align}
Here ``polar'' or ``scalar type'' describes the angular dependence.
It does not mean the extra spin-zero particle of quadratic gravity.
No amplitude in \eqref{eq:seven} has been set to zero by a gauge
choice. The method is the Kodama--Ishibashi construction
\cite{Kodama:2003jz}, with conventions related below to the usual
four-dimensional metric amplitudes
\cite{Regge:1957td,Zerilli:1970se,Moncrief:1974am}.

\subsection{The Kodama--Ishibashi gauge invariants}\label{sec:KIdefinitions}
KI use the normalized harmonics
\begin{equation}
 S_A=-\frac1kY_A,\qquad S_{AB}=\frac1{k^2}Y_{AB},
\end{equation}
and write
\begin{equation}
 h_{ab}=f_{ab}Y,\qquad h_{aA}=rf_aS_A,\qquad
 h_{AB}=2r^2(H_L\gamma_{AB}Y+H_TS_{AB}).
\end{equation}
Comparison with \eqref{eq:seven} gives
\begin{align}\label{eq:KIdictionary}
 f_{tt}&=AH_0,\quad f_{tr}=H_1,\quad f_{rr}=H_2/B,\nonumber\\
 f_t&=-kj_0/r,\quad f_r=-kJ/r,\quad
 H_L=K/2,\quad H_T=k^2G_{\rm met}.
\end{align}
The factors of $k$ and $r$ are part of the harmonic convention.
They are needed when comparing equations written in different bases.

The compensating covector and invariants are
\begin{align}\label{eq:KIkinematic}
 X_a&=\frac r k\left(f_a+\frac r kD_aH_T\right),\\
 F_{ab}&=f_{ab}+D_aX_b+D_bX_a,\qquad
 F=H_L+\frac12H_T+\frac{D^ar}{r}X_a.
\end{align}
Their gauge changes cancel. The cancellation uses the spherical
geometry, not the Einstein field equations. These definitions remain
valid when $A\ne B$.

To display them explicitly, introduce
\begin{equation}
 p_t=j_0,\qquad p_r=J-r^2G_{\rm met}',\qquad u=A'/A.
\end{equation}
Then $X_a=-p_a$, and \eqref{eq:KIkinematic} becomes
\begin{align}\label{eq:invariants}
 U&=H_0+Bup_r,&V&=H_2-2Bp_r'-B'p_r,\nonumber\\
 W&=K+6G_{\rm met}-\frac{2B}{r}p_r,&
 S&=H_1-j_0'+uj_0,\\
 F_{tt}&=AU,&F_{rr}&=V/B,&F_{tr}&=S,&2F&=W.\nonumber
\end{align}
For example, a radial gauge displacement changes $p_r$ by $-F_r$
and $H_0$ by $BuF_r$. Their changes cancel in $U$. The other
cancellations follow in the same way; all gauge transformations are
listed in Appendix~\ref{app:operator}.

The seven amplitudes have three stationary gauge functions, leaving
four invariants. Under $t\mapsto-t$, $S$ is odd while $U,V,W$ are
even. We study the even block. Thus $S=0$ selects the symmetry of
the applied static electric tide; it is not a coordinate gauge choice.
One useful invariant representative of this block is
\begin{equation}\label{eq:GIrepresentative}
 H^{\GI}_{tt}=AUY,\quad H^{\GI}_{rr}=VY/B,\quad
 H^{\GI}_{AB}=r^2W\gamma_{AB}Y.
\end{equation}
The original metric is recovered by adding back the arbitrary gauge
functions. We use \eqref{eq:GIrepresentative} to evaluate the field
equations, not to discard equations before varying the action.

\subsection{The scalar-curvature equation and its boundary conditions}
The trace perturbation $\phi=\delta R$ is gauge invariant because
$\bar R=0$. Linearizing \eqref{eq:trace} gives
\begin{equation}\label{eq:scalar}
 (\bBox-m_0^2)\phi=0.
\end{equation}
The distinction between the angular scalar type and a physical
scalar field is familiar from gravitational perturbations in $f(R)$
models \cite{Bhattacharyya:2017tyc,Bhattacharyya:2018qbe}.
Before imposing boundary conditions, the scalar part can be separated
by the exact linear transformation
\begin{equation}\label{eq:hhat}
 \widehat h_{\mu\nu}=h_{\mu\nu}+2\beta\phi\bg_{\mu\nu},\qquad
 \delta E^{\rm EW}[\widehat h]=0,\qquad\delta R[\widehat h]=0.
\end{equation}
EW denotes the Einstein--Weyl part of the field equation. This
transformation separates the scalar equation; it does not separate
the two spin-two sectors on a non-Einstein background.

For $\phi=\Phi(r)Y$, the radial equation is
\begin{equation}\label{eq:scalar_radial_main}
 (r^2\sqrt{AB}\,\Phi')'
 -r^2\sqrt{A/B}\left(m_0^2+\frac6{r^2}\right)\Phi=0.
\end{equation}
Multiply by $\Phi^*$ and integrate over the exterior. The boundary
term vanishes at a regular nonextremal horizon. It also vanishes at
infinity when there is no applied scalar tide and the scalar decays.
For $m_0^2>0$, the remaining integral is a sum of nonnegative terms,
so
\begin{equation}\label{eq:scalarzero}
 \Phi=0,\qquad h=\widehat h.
\end{equation}
Both boundary conditions are needed. Horizon regularity alone does
not require the scalar amplitude to vanish. Appendix~\ref{app:scalar}
gives the full argument. When $\beta=0$, the trace instead imposes
$\delta R=0$ directly.

On $\bar R=0$, every term in the linearized equation that comes
from $\beta R^2$ is proportional to $\phi$ or its derivatives:
\begin{equation}
 \delta E^{(\beta)}_{\mu\nu}
 =2\beta[\bar R_{\mu\nu}\phi+\bg_{\mu\nu}\bBox\phi
                                  -\bn_\mu\bn_\nu\phi].
\end{equation}
It therefore vanishes for the boundary conditions used here. The
background and gravitational response are independent of the
nonnegative value of $\beta$, at fixed $\alpha,r_h$ and branch.
This conclusion does not cover an applied scalar tide, matter, or
$\beta<0$.

\subsection{\texorpdfstring{$P_2$, $\eta$ and $\Gamma$}{P2, eta and Gamma}}
The original metric equations are fourth order. To write them as a
larger system of lower-order equations, we introduce the perturbation
of the Ricci tensor. It is not an independent test field: its
definition in terms of the metric is kept as an equation.

On a non-Einstein background, raw $\delta R_{\mu\nu}$ is gauge
dependent. If $\mathsf P$ is the vector built from the KI gauge
compensator, the invariant tensor is
\begin{equation}\label{eq:qGI}
 q^{\GI}_{\mu\nu}=\delta R_{\mu\nu}-\Lie_{\mathsf P}\bar R_{\mu\nu}
                       =\delta\operatorname{Ric}_{\mu\nu}[H^{\GI}].
\end{equation}
We denote this tensor by $q$ below. Its time-even polar components
have the harmonic expansion
\begin{align}\label{eq:Ricciharmonics}
 q_{tt}&=AQ_0Y,&q_{rr}&=\frac{P_2}{B}Y,&q_{rA}&=\eta Y_A,\nonumber\\
 q_{AB}&=r^2(Q_K\gamma_{AB}Y+2\Gamma Y_{AB}).
\end{align}
This equation defines the new radial variables. In particular,
$P_2$ is the radial--radial Ricci amplitude, $\eta$ is the
radial--angular amplitude, and $\Gamma$ is the trace-free angular
amplitude. $\Gamma$ is not the metric amplitude $G_{\rm met}$, and
none of these quantities is a new gauge function.

They can be calculated directly from $U,V,W$. With $v=B'/B$,
three useful component relations are
\begin{align}\label{eq:Ricciexplicit}
 \Gamma={}&\frac{U-V}{4r^2},\\
 \eta={}&\frac12(U'-W')+\frac{ru-2}{4r}U+\frac{ru+2}{4r}V,\\
 P_2={}&\frac B2U''-BW''+\frac{B(2u+v)}4U'
       +\frac{B(ru+4)}{4r}V'
       -\frac{B(rv+4)}{2r}W'+\frac3{r^2}V.
\end{align}
Thus introducing $q$ moves some metric derivatives into named
variables; it does not change the solution space when the defining
relations are enforced. The angular relation already gives
\begin{equation}\label{eq:VfromGamma}
 V=U-4r^2\Gamma.
\end{equation}
The Einstein relation $V=U$ is recovered only when this Ricci
amplitude vanishes. Assuming it on the hairy background would remove
an allowed gravitational channel.

The scalar condition is also more than a trace-free condition on
$q$ alone. Since the inverse metric changes, $\delta R=0$ reads
\begin{equation}\label{eq:RicciTraceMain}
 -Q_0+P_2+2Q_K+\rho_tU-\rho_rV-2\rho_oW=0,
\end{equation}
where $\rho_t=\bar R^t{}_t$, $\rho_r=\bar R^r{}_r$ and
$\rho_o=\bar R^\theta{}_\theta=\bar R^\varphi{}_\varphi$. The linearized
Bianchi identity likewise contains background Ricci terms. We keep
these terms when reducing the equations.

\subsection{Reduction to six first-order equations}\label{sec:sixstate}
The reduction uses three sets of relations: the Ricci definitions,
the scalar and Bianchi constraints, and the linearized field
equations. We first use \eqref{eq:VfromGamma} to remove $V$.
The remaining Ricci definitions and constraints determine the
dependent components $Q_0,Q_K$ and give first derivatives for the
remaining variables. A nonredundant set of field equations supplies
the curvature evolution. The result is
\begin{equation}\label{eq:state}
 \bm s'=L(r)\bm s,\qquad
 \bm s=(U,W,P_2,\eta,\Gamma,\Gamma')^T.
\end{equation}
The complete entries of $L$ are given in Appendix~\ref{app:L}.
They are functions of the numerical background and its derivatives.
No unspecified potential needs to be fitted.

Two rows help show what this matrix represents. The definition of
$\eta$ gives
\begin{equation}\label{eq:Wprimeexample}
 W'=U'+uU-2\eta-2r(ru+2)\Gamma.
\end{equation}
The angular trace-free field equation gives
\begin{align}\label{eq:Gammaprimeexample}
 \Gamma''={}&\frac{Bu-B'}{2Br^3}U-\frac{2}{r^3}\eta
 +\frac{-2Bru-2B+m_2^2r^2+6}{Br^2}\Gamma
 -\frac{Bru+4B+rB'}{2Br}\Gamma'.
\end{align}
For example, eliminating the dependent Ricci components gives the
first metric row explicitly:
\begin{align}\label{eq:Uprimeexample}
 U'={}&-\frac{Br^2u^2+2Bru+4}{Br^2u}U
       +\frac{4}{Br^2u}W+\frac{2}{Bu}P_2\nonumber\\
 &+\frac{2(ru+2)}{ru}\eta
       +\frac{2(Br^2u^2+4Bru+4B-4)}{Bu}\Gamma.
\end{align}
Substituting this into \eqref{eq:Wprimeexample} determines $W'$.
The remaining independent curvature equations and Bianchi constraints
give $P_2'$ and $\eta'$. Together with the identity
$\Gamma'=s_6$ and \eqref{eq:Gammaprimeexample}, these relations
give the six rows of \eqref{eq:state}. This is why the enlarged
state is useful: it gives a closed propagation system while keeping
the Ricci definitions and constraints. We check the reduction by
substituting the reconstructed solution into the original equations.
The covariant fourth-order operator and all component maps are in
Appendix~\ref{app:operator}.

\subsection{Keeping the static KI variable in the reduced system}
KI's frequency-domain definition cannot be evaluated by dividing
by $\omega$ and then setting $\omega=0$. We use their static
combination instead \cite{Kodama:2003jz}. Choose a fixed nonzero
reference mass $M_{\rm ref}$ and define
\begin{equation}
 x=2M_{\rm ref}/r,\quad f_{\rm ref}=1-x,\quad H=4+3x,
 \quad Q_X=18x^2+28x-32,\quad Q_Y=6x^2-20x.
\end{equation}
Then
\begin{equation}\label{eq:KI}
 \Psi=\frac{-Q_YU-Q_XV+(Q_X-Q_Y)W}{12f_{\rm ref}'H}
       =c_0(r)\bm s.
\end{equation}
On Schwarzschild this is the static KI variable. On the two-function
hairy background it defines a particular continuation of that
invariant combination. The actual $A,B$ are never replaced by
$f_{\rm ref}$. The reference mass specifies the variable, while the
physical mass used in the far-field tidal basis is determined from
the background. Changing the reference basis must not change the
Love number extracted from the physical metric.

Choose three fields and three complementary state components,
\begin{equation}
 \bm y=(\Psi,\eta,\Gamma)^T,\qquad
 \bm w=(W,P_2,\Gamma')^T,\qquad
 \begin{pmatrix}\bm y\\\bm w\end{pmatrix}=T(r)\bm s.
\end{equation}
Differentiating this definition and using \eqref{eq:state} gives
\begin{equation}\label{eq:blocksystem}
 \begin{pmatrix}\bm y'\\\bm w'\end{pmatrix}
 =\begin{pmatrix}\mathsf C&\mathsf M\\\mathsf N&\mathsf F\end{pmatrix}
  \begin{pmatrix}\bm y\\\bm w\end{pmatrix},\qquad
 \begin{pmatrix}\mathsf C&\mathsf M\\\mathsf N&\mathsf F\end{pmatrix}
 =(T'+TL)T^{-1}.
\end{equation}
Where $\mathsf M$ is invertible, the first row determines
$\bm w=\mathsf M^{-1}(\bm y'-\mathsf C\bm y)$. Differentiating
once more and using the second row removes $\bm w$ entirely:
\begin{align}\label{eq:three}
 \bm y''+P\bm y'+Q\bm y&=0,\\
 \mathsf J&=\mathsf C\mathsf M+\mathsf M'+\mathsf M\mathsf F,
 \qquad\mathsf K=\mathsf C'+\mathsf C^2+\mathsf M\mathsf N,\nonumber\\
 P&=-\mathsf J\mathsf M^{-1},\qquad
 Q=\mathsf J\mathsf M^{-1}\mathsf C-\mathsf K.\nonumber
\end{align}
These are the three coupled second-order radial equations. In
particular,
\begin{equation}\label{eq:KIsourced}
 \Psi''+P_{11}\Psi'+Q_{11}\Psi
 =-P_{12}\eta'-P_{13}\Gamma'-Q_{12}\eta-Q_{13}\Gamma.
\end{equation}
The KI variable is generally driven by the two retained curvature
amplitudes. It is not a closed massless equation on the hairy
background.

\subsection{The three Schr\"odinger-like equations}
Introduce the tortoise coordinate and a positive exterior function
$c(r)$,
\begin{equation}
 \frac{\dd r_*}{\dd r}=\frac1c,\qquad c=\sqrt{AB},\qquad
 \dot{\ }=\frac{\dd}{\dd r_*}.
\end{equation}
Equation~\eqref{eq:three} becomes
\begin{equation}
 \ddot{\bm y}+P_*\dot{\bm y}+Q_*\bm y=0,
 \qquad P_*=cP-c'\Id,\quad Q_*=c^2Q.
\end{equation}
Let $\bm y=S_{\rm tr}\bm\Upsilon$, where the matrix $S_{\rm tr}$
solves $\dot S_{\rm tr}=-P_*S_{\rm tr}/2$. This removes the first
derivative and gives
\begin{equation}\label{eq:matrixSch}
 \ddot{\bm\Upsilon}-\mathcal V\bm\Upsilon=0,\qquad
 \mathcal V=S_{\rm tr}^{-1}
 \left(\tfrac12\dot P_*+\tfrac14P_*^2-Q_*\right)S_{\rm tr}.
\end{equation}
In components,
$\ddot\Upsilon_i-\sum_{j=1}^3\mathcal V_{ij}\Upsilon_j=0$.
The off-diagonal entries describe the coupling. The matrix need
not be symmetric in this field basis. Also, $\Upsilon_1$ need not
equal $\Psi$, because $S_{\rm tr}$ mixes the fields.

There are six local integration constants in the selected static
even sector, just as in \eqref{eq:state}. This count does not give
the number of all time-dependent polarizations of quadratic gravity.
The equations were derived in the static limit. Adding
$\omega^2\Id$ to them would require a separate time-dependent
derivation. For the global numerical calculation we use
\eqref{eq:state}; a particular three-field change of variables can
become singular at a regular radius even when that state system
remains usable. Appendix~\ref{app:operator} explains the treatment
of these apparent singularities.

\subsection{Why simple attempts to separate the equations fail}\label{sec:decoupling}
Diagonalizing $\mathcal V$ at each radius does not remove the
coupling. Its eigenvectors change with radius, and their derivatives
produce new off-diagonal terms. The simple curvature shift that
separates the Schwarzschild massless field also fails when the
background Ricci tensor is nonzero.

We tested more general transformations in restricted classes.
Tests on exact local background data rule out the chosen point
transformations, ordinary Darboux maps to specified target potentials,
and a finite rational ansatz for solution-space projectors
\cite{Glampedakis:2017rar,Samsonov:2004ri}. These tests do not prove
that every possible separation is impossible on a global L\"u
solution.

A more general local transformation can be constructed by integrating
the full fundamental matrix. It can even map the system to free
equations. However, the original propagation and boundary information
then sit in the transformation itself. The target fields do not
automatically describe separate massless and massive tides. Such a
map cannot be used to infer a zero metric response. The detailed
tests and construction are in Appendix~\ref{app:decoupling}.

\subsection{The Schwarzschild result}\label{sec:Schwarzschild}
The Ricci-flat background is special. After scalar separation,
$q$ is transverse and traceless, and
\begin{equation}\label{eq:Schsplit}
 (\Delta_L+m_2^2)q=0,\qquad
 h^{(0)}=h+\frac{2}{m_2^2}q,\qquad
 h^{(m)}=-\frac{2}{m_2^2}q,
 \qquad\delta\operatorname{Ric}[h^{(0)}]=0,
\end{equation}
where $\Delta_Lq=-\bBox q-2\bar R_{\mu\rho\nu\sigma}q^{\rho\sigma}$.
The field $h^{(0)}$ therefore obeys the ordinary massless Einstein
equation. This separation agrees with the massless and massive
sectors used in perturbative studies of Schwarzschild
\cite{Brito:2013wya,Antoniou:2024jku,Knoska:2025jdj}.

The regular unit-source quadrupole is
\begin{equation}\label{eq:Schpoly}
 U=V=r(r-2M),\qquad W=r^2-2M^2,\qquad
 h_{tt}/Y=(r-2M)^2.
\end{equation}
There is no independent $r^{-3}$ massless response. The other
massless solution is singular at the horizon. A permitted decaying
massive solution has an exponential tail, so it does not add an
independent massless power law at $m_2r\gg1$. Thus, for a regular
horizon and only a massless applied quadrupole,
\begin{equation}\label{eq:Schzero}
 \lambda_E^{\rm Schwarzschild}=k_2^{\rm Schwarzschild}=0.
\end{equation}
This is the Schwarzschild result in the full quadratic theory for
the stated tidal experiment. It does not say that massive modes
cannot propagate or that all massive boundary problems have a
unique solution. Appendix~\ref{app:Schwarzschild} verifies the
radial equation and regularity in ingoing coordinates.

\section{Calculating the tidal Love number}\label{sec:TLN}
\subsection{What is applied and what is measured}\label{sec:normalization}
We apply a static massless electric quadrupolar tide. We require
a regular horizon and no independently applied scalar or massive
tensor tide. The decaying massive tensor fields are allowed to
respond. They must not be set to zero throughout the exterior:
on the hairy background the gravitational channels are coupled.

The external tide grows with distance from the black hole. The
perturbed spacetime is therefore not globally asymptotically flat.
As in the usual tidal problem, its large-radius expansion describes
a buffer region: far from the black hole, but still much closer
to it than to the distant source of the tide
\cite{Binnington:2009bb,Damour:2009vw,Porto:2016pyg}.
In the original metric, fix the multipole convention by
\begin{equation}\label{eq:multipoles}
 g_{tt}=-1+\frac{2M}{r}-\mathcal E_{ij}x^ix^j
             +\frac{3Q_{ij}n^in^j}{r^3}+\cdots,
 \qquad Q_{ij}=-\lambda_E\mathcal E_{ij}.
\end{equation}
Here $\mathcal E_{ij}$ is the applied tidal tensor and $Q_{ij}$
is the induced mass quadrupole. Both are symmetric and trace free.
The omitted terms include subleading terms from the applied
source and the massive responses.

For one harmonic, let $a$ multiply the massless source solution
and $b$ the massless response solution. Normalize their leading
$h_{tt}/Y$ terms to $r^2$ and $r^{-3}$. Then
$a=-\mathcal E_m$, $b=3Q_m$, and
\begin{equation}\label{eq:love}
 \lambda_E=\frac{b}{3a},\qquad
 k_2^{(r_h)}=\frac{3\lambda_E}{2r_h^5}
                    =\frac{b}{2ar_h^5},\qquad
 \Lambda_E=\frac{\lambda_E}{M^5}.
\end{equation}
We use $r_h$ to make $k_2$ dimensionless. This choice must be stated
because $r_h\ne2M$ for the non-Schwarzschild solution. Normalizing
by $2M$ changes the reported dimensionless number, but not
$\lambda_E$ or the physical quadrupole.

\subsection{Building the horizon and large-radius solutions}
Equation~\eqref{eq:state} has six local integration constants.
The horizon condition and the applied tide select one solution
up to its overall amplitude. We construct it from both ends
instead of integrating one arbitrary solution outward and fitting
its tail.

Near the horizon, expand the six-state matrix in powers of
$x=r-r_h$,
\begin{equation}\label{eq:horLaurentMain}
 L=\frac{R_{\rm hor}}{x}+L_0+L_1x+\cdots.
\end{equation}
For the reference solution the indicial polynomial is
$\rho^3(\rho+1)^3$. Three independent solutions have regular
analytic horizon series. Choose their free leading data as
$(W_h,P_{2h},\Gamma_h)=(1,0,0),(0,1,0),(0,0,1)$.
The recursion determines the higher coefficients of each column.
Evaluating these series at $r_h+\epsilon_h$ gives a six-by-three
matrix $H$ of regular solutions. We propagate its columns outward
to a matching radius $r_m$.

Horizon regularity is a tensor condition. In the ingoing coordinate
$v=t+r_*$, the invariant representative has
\begin{equation}\label{eq:EFmain}
 h_{vv}/Y=AU,\qquad h_{vr}/Y=-\sqrt{A/B}\,U,\qquad
 h_{rr}^{\rm EF}/Y=(U+V)/B.
\end{equation}
The regular series makes these combinations finite. Finiteness of
a single master variable in Schwarzschild coordinates would not
be an adequate check.

At large radius there are six asymptotic columns in this static
block. Two have massless power-law behavior: the applied tide
$T$ and response $R$. There are also two decaying massive columns
$D_1,D_2$ and two growing massive columns $G_1,G_2$. The two
columns for each massive sign are independent radial solutions
within the selected symmetry block, not a count of all massive
spin-two polarizations.

We exclude $G_1,G_2$ because there is no applied massive tide.
We retain $D_1,D_2$ because they describe induced massive fields.
The allowed large-radius solution is therefore
\begin{equation}\label{eq:outercolumnsMain}
 \bm s=aT+bR+d_1D_1+d_2D_2.
\end{equation}
We initialize these four columns at an outer radius $R_o$ and
propagate them inward to $r_m$. The background mass entering
their expansion is the asymptotic mass of the numerical solution,
not the auxiliary reference mass used in the KI definition.

The complete massless source is important. In the asymptotically
Schwarzschild region, it is
\begin{equation}\label{eq:sourcefullMain}
 U_T=V_T=r(r-2M),\qquad W_T=r^2-2M^2,\qquad
 h_{tt}^T/Y=r^2-4Mr+4M^2.
\end{equation}
The $Mr$ and $M^2$ terms belong to the applied tide. They must
not be mistaken for a response. We use a separate response
solution normalized to $h_{tt}^R/Y=r^{-3}+O(r^{-4})$.
The massive columns contain $e^{\pm m_2r}$ times radial powers
and inverse-radius series. Their explicit construction is given
in Appendix~\ref{app:matching}.

The endpoint expansions use the Schwarzschild asymptotic operator,
while propagation uses the full numerical hairy operator. At a
finite $R_o$, this neglects exponentially small background-hair
corrections to the endpoint basis. Moving $R_o$ outward is therefore
an essential test. High precision alone does not remove this error.

\subsection{Matching the two solutions}\label{sec:matching_main}
At $r_m$, a regular horizon solution is $H\bm c$, where
$\bm c$ has three components. Set the applied amplitude to $a=1$.
Continuity of the six-state solution gives
\begin{equation}\label{eq:matchingMain}
 H(r_m)\bm c=T(r_m)+bR(r_m)+d_1D_1(r_m)+d_2D_2(r_m).
\end{equation}
Equivalently,
\begin{equation}\label{eq:matchingMatrixMain}
 \big[H,-R,-D_1,-D_2\big]_{r_m}
 \begin{pmatrix}c_1\\c_2\\c_3\\b\\d_1\\d_2\end{pmatrix}=T(r_m).
\end{equation}
There are six equations and six unknowns. The first three unknowns
choose the regular horizon combination. The last three are the
massless and massive response amplitudes. No response amplitude
is fixed in advance. If the matching matrix is nonsingular,
the imposed source determines them all.

\begin{figure}[tbp]
 \centering\begin{tikzpicture}[
 font=\small,node distance=7mm,
 box/.style={draw=black!65,rounded corners=2pt,fill=blue!3,
   align=center,text width=6.0cm,minimum height=10mm,inner sep=4pt},
 central/.style={box,text width=10.4cm},
 flow/.style={-{Stealth[length=2mm]},thick,draw=black!70}]
 \node[central] (bg)
 {Use the numerical background and the six-state equation $\bm s'=L\bm s$};
 \node[box,below left=9mm and 3mm of bg.south] (horizon)
 {Horizon: construct three regular series\\with independent $(W_h,P_{2h},\Gamma_h)$};
 \node[box,below right=9mm and 3mm of bg.south] (outer)
 {At $R_o$: construct $T,R,D_1,D_2$\\Set the growing massive sources to zero};
 \node[box,below=of horizon] (hout)
 {Propagate the three columns $H$ outward\\to the matching radius $r_m$};
 \node[box,below=of outer] (oin)
 {Propagate the four allowed columns inward\\from $R_o$ to $r_m$};
 \coordinate (joinlevel) at (bg.center |- oin.south);
 \node[central,below=10mm of joinlevel] (match)
 {Set $a=1$ and solve six matching equations:\\
 $H\bm c=T+bR+d_1D_1+d_2D_2$};
 \node[central,below=of match] (reconstruct)
 {Reconstruct the physical metric and check the original equations};
 \node[central,below=of reconstruct,fill=green!5] (love)
 {Read the massless ratio $b/a$ and calculate\\
 $\lambda_E=b/(3a)$,\qquad $k_2^{(r_h)}=b/(2ar_h^5)$};
 \node[central,below=of love] (accuracy)
 {Vary $R_o$, $r_m$, the horizon offset, series orders and precision};
 \draw[flow] (bg.south)-|(horizon.north);
 \draw[flow] (bg.south)-|(outer.north);
 \draw[flow] (horizon)--(hout);
 \draw[flow] (outer)--(oin);
 \draw[flow] (hout.south)--([xshift=-3.3cm]match.north);
 \draw[flow] (oin.south)--([xshift=3.3cm]match.north);
 \draw[flow] (match)--(reconstruct);
 \draw[flow] (reconstruct)--(love);
 \draw[flow] (love)--(accuracy);
\end{tikzpicture}
 \caption{The Love-number calculation. Three horizon-regular columns
 are propagated outward and four allowed asymptotic columns inward.
 A single six-by-six matching solve fixes the unit-source solution.
 The physical metric is then reconstructed before the source and
 response coefficients are converted to a Love number.}
 \label{fig:love_algorithm}
\end{figure}
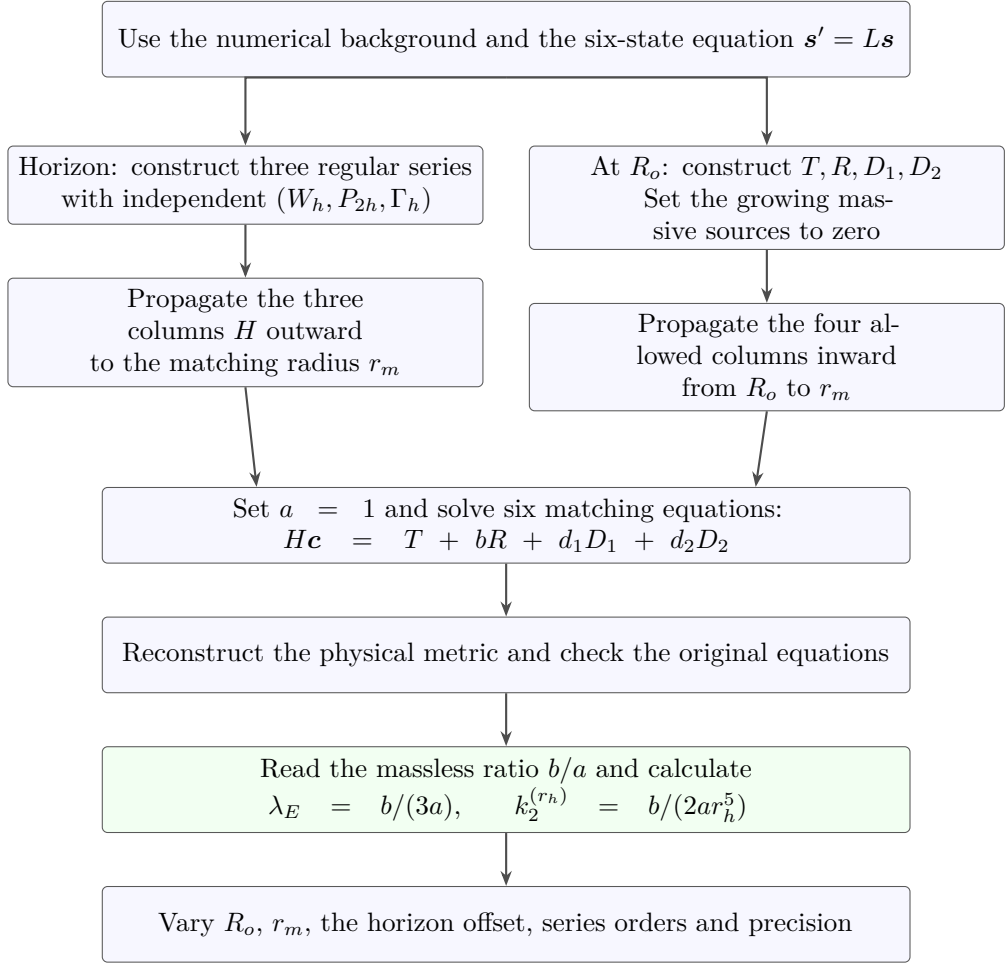

This construction also explains why a negative Love number is not
put in by hand. Its sign is the sign of the coefficient $b$ selected
by the global horizon and source conditions. The matrix can become
ill conditioned near a homogeneous static mode. A possible pole
in the response must therefore be tested using the matching matrix,
not inferred from a steep curve alone. Figure~\ref{fig:love_algorithm}
summarizes the calculation.

After matching, we reconstruct $V$ from \eqref{eq:VfromGamma}
and the remaining Ricci components from their definitions.
We substitute this solution into the original metric equations,
the Ricci definitions and the Bianchi constraints. We also vary
the horizon offset, expansion orders, matching radius and outer
radius. A small linear-solve residual only checks the solve in
\eqref{eq:matchingMatrixMain}; it does not estimate the uncertainty
from the finite outer boundary.

\subsection{The reference result and checks of accuracy}
For $\ahat=1/2$, the refined calculation uses $R_o/r_h=40$,
$r_m/r_h=6$ and $\epsilon_h/r_h=0.0004$. The horizon expansion
has order 16, the massive asymptotic series uses 28 terms, and
the propagation uses high-precision Taylor integration. The result is
\begin{align}\label{eq:mainresult}
 b/(ar_h^5)&\simeq-0.3674783,&
 \lambda_E/r_h^5&\simeq-0.1224928,\nonumber\\
 k_2^{(r_h)}&\simeq-0.1837391,&
 \Lambda_E&\simeq-77.1305.
\end{align}
For the same deformability, normalization by $2M$ gives
$k_2^{(2M)}\simeq-3.61549$.

\begin{table}[tbp]
\centering
\caption{Sensitivity to the outer endpoint in the reference tidal
calculation. The same refined background is used in all rows.
The differences measure finite-endpoint sensitivity; they are not
statistical error bars.}\label{tab:cutoff}
\begin{tabular}{@{}rrr@{}}\toprule
$R_o/r_h$ & $b/(ar_h^5)$ & $k_2^{(r_h)}$\\\midrule
28 & $-0.3677104436$ & $-0.1838552218$\\
36 & $-0.3674785754$ & $-0.1837392877$\\
40 & $-0.3674782548$ & $-0.1837391274$\\\bottomrule
\end{tabular}
\end{table}

The final change in $k_2^{(r_h)}$ is about $1.60\times10^{-7}$.
The corresponding change in $\Lambda_E$ is $6.73\times10^{-5}$.
We do not fit an infinite-radius extrapolation from these three
points. The number of printed digits is set by these changes,
not by the number of working digits in the integration.
The original equations and constraints also have small residuals,
but those residuals test a different source of error.

An additional check uses the Weyl curvature in a background
principal tetrad. In our convention,
\begin{equation}\label{eq:curvaturecheck}
 \delta\Psi_0=\frac{\sqrt{24}}{4Ar^2}(U+V)\,{}_2Y_{2m},\qquad
 \Xi=\frac{2A}{\sqrt{24}}\frac{\delta\Psi_0}{{}_2Y_{2m}}
       =\frac{U+V}{2r^2}.
\end{equation}
The normalized source has $\Xi_T=1-2M/r$, and the response begins
at $r^{-5}$. Subtracting the full source gives the same $b/a$ as
the metric calculation. This checks the normalization using a
different curvature quantity. It is not a new scattering-amplitude
or worldline matching calculation.

\begin{table}[tbp]
\centering
\caption{The reference point and two nearby validated calculations.
The last column gives the final outer-endpoint change.}
\label{tab:nearby}
\begin{tabular}{@{}rrrr@{}}\toprule
$\ahat$ & $M/r_h$ & $k_2^{(r_h)}$ & $|\Delta k_2|$\\\midrule
0.49 & 0.2566565674 & $-0.2181892855$ & $1.39\times10^{-7}$\\
0.50 & 0.2755350525 & $-0.1837391274$ & $1.60\times10^{-7}$\\
0.51 & 0.2937738779 & $-0.1535772784$ & $1.37\times10^{-7}$\\\bottomrule
\end{tabular}
\end{table}

The neighboring points in Table~\ref{tab:nearby} support the local
coupling trend. The same horizon-series method reproduces the
Schwarzschild polynomial and its zero response. Appendix~\ref{app:matching}
gives further numerical details.

\subsection{Changing the coupling at fixed horizon radius}\label{sec:scan}
\begin{figure}[tbp]
 \centering\includegraphics[width=\linewidth]{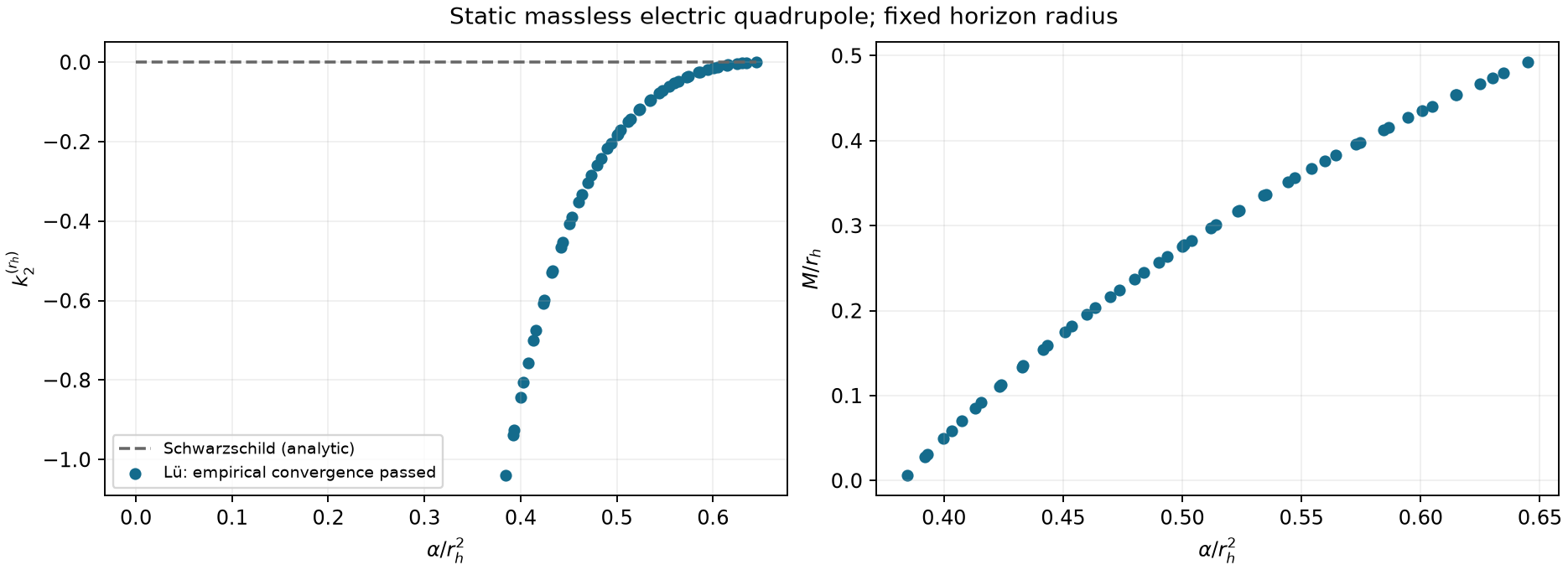}
 \caption{The fixed-horizon scan of the selected non-Schwarzschild
 branch. Left: the horizon-normalized Love number, compared with
 the analytic Schwarzschild value. Right: the changing asymptotic
 mass. The marker label refers to the scan's empirical convergence
 criterion. Quantitative endpoint checks are listed for the reference
 and neighboring calculations in Tables~\ref{tab:cutoff} and
 \ref{tab:nearby}; no error bars for the entire curve are inferred
 from these three calculations.}\label{fig:scan}
\end{figure}

The response is negative on the sampled branch. Its magnitude
decreases as $\ahat$ increases, while $M/r_h$ increases toward
$1/2$. This is consistent with approaching the Schwarzschild
intersection, provided the static matching problem remains regular
in that limit. The scan ends at $\ahat=0.645$, short of the
expected intersection. It does not establish an
$\alpha\to\infty$ limit or exclude new behavior on another
continuation of the branch.

At the other end, decreasing $\ahat$ increases $p=m_2r_h$.
The mass of this selected family decreases toward zero near
$\ahat\simeq0.383$. This is not the Einstein limit of a
fixed-mass black hole. The Schwarzschild branch exists separately
down to $\alpha=0$, while this positive-mass non-Schwarzschild
segment does not reach that value. Its Love number therefore
need not approach the Schwarzschild zero at the left edge of
the plot.

The steep negative trend near the low-mass end does not by itself
prove a divergence or a static response pole. More closely spaced
points and their matching diagnostics would be needed. Because
$k_2^{(r_h)}$ uses the fixed horizon radius, the trend is not
caused simply by division by a small mass. Mass normalization
would amplify the effect further:
\begin{equation}\label{eq:massnorm}
 \Lambda_E=\frac23k_2^{(r_h)}\left(\frac{r_h}{M}\right)^5.
\end{equation}

Keeping $p$ fixed and changing $\alpha$ is a different operation.
It rescales the same dimensionless solution:
\begin{equation}\label{eq:scaling}
 r_h,M\propto\sqrt\alpha,\qquad
 \lambda_E\propto\alpha^{5/2},\qquad
 k_2^{(r_h)},\Lambda_E=\text{constant}.
\end{equation}
Conversely, a fixed-mass scan follows a different path, with
$\alpha/M^2=\ahat/(M/r_h)^2$. These choices answer different
questions; neither should be substituted for the other without
changing the interpretation of the plot.

\subsection{The meaning of a negative response and the EFT question}
In \eqref{eq:multipoles}, a negative $\lambda_E$ reverses the
sign of the induced quadrupole relative to a positive-deformability
object under the same tidal convention. This is a statement about
the conservative static response. It is not a dissipative phase
shift or, by itself, a stability criterion.

In particular, the usual Gregory--Laflamme-type instability is a
time-dependent monopole problem, whereas the coefficient calculated
here belongs to a static quadrupole. The two sectors have different
angular momentum and different boundary conditions
\cite{Brito:2013wya,Held:2022abx,East:2023nsk}.
A negative $\ell=2$ response does not establish an unstable
$\ell=0$ mode. Nor do stable test-field or axial spectra prove
stability of every polar gravitational mode
\cite{Cai:2015fia,Zinhailo:2018ska,Antoniou:2024jku,Konoplya:2025afm}.

Finally, the metric coefficient must still be related to a
consistently defined observable in an effective theory. For example,
with the simple response convention
\begin{equation}
 L_Q=-\frac{Q_{ij}Q^{ij}}{4\lambda_E}
                     -\frac12Q_{ij}\mathcal E^{ij},
\end{equation}
eliminating $Q_{ij}$ gives $L_{\rm tide}=\lambda_E\mathcal E^2/4$.
This algebra fixes a convention; it does not perform the full
matching. In higher-curvature gravity, bulk operators and induced
worldline terms must be treated together
\cite{Porto:2016pyg,Kulkarni:2024ghc,Wang:2026qst}.
An independent scattering or worldline calculation is needed to
complete that step for this non-Schwarzschild branch
\cite{Ivanov:2022qqt}.

\section{Discussion and conclusions}\label{sec:conclusions}
We have studied the static gravitational quadrupole of two
black-hole branches in the same quadratic theory. The Schwarzschild
branch has zero massless Love number for a regular horizon and no
applied massive or scalar tide. On the non-Schwarzschild branch,
the background Ricci tensor couples the gravitational channels.
The resulting metric response is nonzero and negative in the
convention defined in this paper. At $\alpha/r_h^2=1/2$,
$k_2^{(r_h)}\simeq-0.1837391$, with a final outer-endpoint
change of about $1.60\times10^{-7}$.

The calculation uses the KI invariants without fixing the metric
gauge. We have defined the curvature amplitudes $P_2$, $\eta$
and $\Gamma$ explicitly and shown how the constrained six-state
system leads to three coupled second-order equations. The scalar
equation separates exactly. Positive scalar mass squared, horizon
regularity and no applied scalar tide then give $\delta R=0$.
The specified gravitational response is consequently independent
of $\beta\geq0$. The same statement does not remove the Ricci
tensor or its spin-two perturbations.

The negative response does not conflict with perturbative
field-redefinition arguments. The low-energy Einstein calculation
and the full non-Schwarzschild solution keep different solution
spaces. An exact change of variables must retain the extra fields
or the full transformed interactions, together with transformed
matter couplings. It cannot change an observable of the same
boundary-value problem. This distinction also helps place the
static result alongside recent time-dependent work on extra
gravitational modes and their suppression near the Einstein limit
\cite{Antoniou:2024jku,Konoplya:2025afm,Antoniou:2026nhh}.

The scan suggests an approach to zero near the Schwarzschild
intersection and a larger negative magnitude toward the low-mass
end. It does not determine a divergence, prove polar dynamical
stability, or exhaust other continuations of the black-hole family.
Further calculations should test these limits, study other
multipoles and frequencies, and compare fixed-horizon and fixed-mass
families. The relation between the metric response and matched
worldline Wilson coefficients remains to be established. Further
work on EFT matching is left to future work.

\section*{Acknowledgments}
The author thanks Gilles Esposito-Far\`ese for helping initiate
this project by asking the questions that shaped its direction.
The author also acknowledges OpenAI's ChatGPT/Codex for assistance
with symbolic calculations, numerical implementation, literature
searches and manuscript preparation. The author is responsible
for the final calculations and interpretation.

\appendix
\section{Background field equations and horizon data}\label{app:background}
We use signature $(-,+,+,+)$ and
$R^\lambda{}_{\rho\mu\nu}=\partial_\mu\Gamma^\lambda_{\nu\rho}
-\partial_\nu\Gamma^\lambda_{\mu\rho}
+\Gamma^\lambda_{\mu\sigma}\Gamma^\sigma_{\nu\rho}
-\Gamma^\lambda_{\nu\sigma}\Gamma^\sigma_{\mu\rho}$.
Varying \eqref{eq:action} gives
\begin{align}\label{eq:full}
 E_{\mu\nu}={}&R_{\mu\nu}-\tfrac12g_{\mu\nu}R
 -2\alpha\Box R_{\mu\nu}-4\alpha R_{\mu\rho\nu\sigma}R^{\rho\sigma}
 \nonumber\\
 &+\tfrac23(\alpha-3\beta)\nabla_\mu\nabla_\nu R
 +\tfrac13(\alpha+6\beta)g_{\mu\nu}\Box R
 +2(\beta+2\alpha/3)RR_{\mu\nu}\nonumber\\
 &+\alpha g_{\mu\nu}R_{\rho\sigma}R^{\rho\sigma}
 -\tfrac12(\beta+2\alpha/3)g_{\mu\nu}R^2=0.
\end{align}
Its trace is \eqref{eq:trace}. Define $u=A'/A$, $v=B'/B$ and the
mixed Ricci eigenvalues
\begin{align}\label{eq:rho}
 \rho_t&=-B\left(\frac{u'}2+\frac{u^2+uv}{4}+\frac ur\right),\\
 \rho_r&=B\left(-\frac{u'}2-\frac{u^2+uv}{4}-\frac vr\right),\\
 \rho_o&=\frac{1-B}{r^2}-\frac{B(u+v)}{2r},
 \qquad\rho_t+\rho_r+2\rho_o=0.
\end{align}
On $\bR=0$ these can equivalently be evaluated as
\begin{equation}
 \rho_t=\frac{B-1+rB'}{r^2},\qquad
 \rho_r=\frac{B-1+rBu}{r^2}.
\end{equation}
For symbolic differentiation, let $\mu=m_2^2$ (a mass squared) and use
\begin{align}\label{eq:backgroundode}
 A''&=\frac{-4A^2B-4A^2rB'+4A^2-4AA'Br-AA'B'r^2+A'^2Br^2}
                    {2ABr^2},\\
 B''&=\frac{\mathcal P_B}{2A^2Br^2(rA'-2A)},\\
 \mathcal P_B={}&-8A^3B^2-4A^3BB'r+4A^3B\mu r^2+8A^3B
 -3A^3B'^2r^2+4A^3B'r-4A^3\mu r^2\nonumber\\
 &-2A^2A'BB'r^2+4A^2A'B\mu r^3
 -3AA'^2B^2r^2-AA'^2BB'r^3+A'^3B^2r^3.
\end{align}
The solved expression for $B''$ is unsuitable at $rA'-2A=0$ unless
its removable on-shell singularity has first been regularized. The
numerical background instead propagates the Ricci eigenvalues. Set
$Z_t=\rho_t'$, $\mathscr S=\rho_t^2+\rho_r^2+2\rho_o^2$,
$\mathscr J=\rho_t\rho_r+(Bu/r)(\rho_r-\rho_o)$ and
$c_b=B'/2+Bu/2+2B/r$. An exterior system is
\begin{align}
 B'&=r\rho_t+(1-B)/r,&
 A'&=Au,& u&=\frac{r\rho_r+(1-B)/r}{B},\\
 \rho_r'&=-\frac u2(\rho_r-\rho_t)-\frac{\rho_t+3\rho_r}{r},&
 \rho_t'&=Z_t,\\
 Z_t'&=\frac{m_2^2\rho_t-2\mathscr J+\mathscr S/2-c_bZ_t
                       +Bu^2(\rho_t-\rho_r)/2}{B}.
\end{align}
This is an exact nonlinear formulation of the $R=0$ background branch,
with the remaining field equation checked as a constraint.

In units $m_2=r_h=1$, the horizon series before normalizing time is
$A=x+a_2x^2+\cdots$, $B=(1+\delta)x+b_2x^2+\cdots$, $x=r-1$.
Substitution in the nonlinear equations determines higher coefficients
from $\delta$. At a finite endpoint $R$, a decaying-hair shooting
condition is
\begin{equation}\label{eq:bgRobin}
 Z_t(R)+\left[1+\frac{1+M_R}{R}\right]\rho_t(R)=0,
 \qquad M_R=\frac R2[1-B(R)].
\end{equation}
This retains the leading logarithmic derivative of
$e^{-r}r^{-1-M}$; it is approximate at finite $R$. The selected nonzero
root is distinguished from the Schwarzschild root during continuation.
For general $m_2$, radii, mass and endpoint series are rescaled using
$z=m_2r$ and $m_2M$; the large-radius logarithmic derivative becomes
$-m_2-(1+m_2M)/r$ at this order. The final constant rescaling of $A$
sets its estimated asymptotic limit to one.

The reference calculation refined the same branch through $R=36$ and
$40$, using high-precision Taylor integration and a horizon series.
For the baseline reference, the recorded settings were an order-22
background series at $x=0.001$, order-30 background propagation and
80-digit arithmetic. A later reproduced scan baseline used 90 working
digits. These are implementation settings; neither working precision
nor the finite-endpoint shooting residual defines an ADM error bar.

\section{What a perturbative field redefinition removes}\label{app:frames}
\subsection{Removing the Ricci-squared terms at first order}
Write $S=S_{\rm EH}+\epsilon S_4$ with the coefficients in
\eqref{eq:redundant}. For a covariant variation $\Delta g$,
\begin{equation}
 \delta(\sqrt{-g}R)=-\sqrt{-g}G^{\mu\nu}\Delta g_{\mu\nu}
                         +\partial_\mu\Theta^\mu.
\end{equation}
In four dimensions $G^\mu{}_\mu=-R$, so the contraction in
\eqref{eq:EHvariation} cancels both Ricci operators. The result is
\begin{equation}\label{eq:EFTequiv}
 S[g(\widetilde g)]
 =S_{\rm EH}[\widetilde g]
 -\frac{\epsilon\alpha}{16\pi G}
      \int\sqrt{-\widetilde g}\,\widetilde{\mathcal G}\,\dd^4x
 +O(\epsilon^2)+S_{\rm boundary}.
\end{equation}
The Euler term and boundary terms require the usual boundary
completion; no claim about entropy or topological observables follows
from discarding them in a local vacuum equation. The cancellation
establishes bulk redundancy at the stated derivative order
\cite{Kamefuchi:1961sb,Deser:1986xr,deRham:2020ejn}.

Keeping the transformed action exactly would retain the higher-order
operators generated by substitution into both $S_{\rm EH}$ and $S_4$.
Truncating \eqref{eq:EFTequiv} and solving Einstein equations exactly
afterward does not give an exact map of the original
fourth-order theory. The redefinition's inverse is itself a derivative
expansion. It is appropriate on the low-momentum solution space where
$|\alpha\nabla^2|,|\beta\nabla^2|\ll1$ and the relevant curvature
scales satisfy the corresponding hierarchy.

For clarity, linear order in the tidal amplitude is not linear order
in $\alpha$. The calculation of the main text takes the former limit
on a background with finite $\ahat$. It does not assume the latter.
Nor does $R=0$ imply $R_{\mu\nu}=0$. On the L\"u branch the Ricci-tensor
part of the redefinition is nonzero, so one cannot infer that a
redefined metric has solved the Einstein equations by setting the
trace to zero.

\subsection{Why the extra solutions are not kept in order reduction}
A schematic linearized scalar-free spin-two operator has the factorized
structure $\Box(\Box-m_2^2)$ on a flat background, with curvature
corrections on a black hole. Its propagator contains two poles:
\begin{equation}
 \frac{1}{k^2(1+k^2/m_2^2)}
       =\frac1{k^2}-\frac1{k^2+m_2^2}.
\end{equation}
For $|k^2|\ll m_2^2$ one expands the heavy propagator as a series of
local terms. That series cannot represent an independent homogeneous
massive solution with $k^2$ of order $m_2^2$. In a static radial problem
the same distinction appears in the solutions $e^{\pm m_2r}$.
The background Ricci equation \eqref{eq:bg} is homogeneous in its
massive part at small hair: a nonzero solution balancing derivatives
against $m_2^2$ has structure on the extra-field scale. A perturbative
order-reduced Einstein solution excludes these data by construction.
This is why a finite-coupling non-Schwarzschild branch is not a
counterexample to the perturbative equivalence statement.

\subsection{The matter couplings must change too}
For a minimally coupled matter sector,
\begin{equation}
 S_m[g(\widetilde g),\psi]
 =S_m[\widetilde g,\psi]
  +\frac\epsilon2\int\sqrt{-\widetilde g}\,
       \widetilde T^{\mu\nu}\Delta g_{\mu\nu}\,\dd^4x
  +O(\epsilon^2).
\end{equation}
Equation~\eqref{eq:worldlineshift} is its point-particle counterpart.
Thus the physical original-frame metric is the composite quantity
$g(\widetilde g)$ in the redefined variables. The external tidal field,
coordinate choice needed to recover areal radius, inferred quadrupole,
and localized operators must be transformed together. An individual
off-shell coefficient need not be invariant under a change of operator
basis. Physical predictions agree once all induced terms and the same
boundary problem are retained.

These observations separate two issues. Bulk operator redundancy
constrains the low-energy Einstein branch; worldline matching fixes the
effective response of a selected object in a specified bulk theory.
The former does not determine all coefficients of the latter for an
additional nonperturbative branch. Conversely a raw metric tail in
one field basis does not by itself complete observable matching.
The distinction is relevant to higher-curvature matching analyses
\cite{Wang:2026qst}. No equality between a new EFT Wilson coefficient
and the number in \eqref{eq:mainresult} is established here.

\section{Why the scalar perturbation vanishes in this tidal problem}\label{app:scalar}
Let $E^{\rm EW}=G_{\mu\nu}-4\alpha B_{\mu\nu}^{\rm Bach}$.
On an on-shell $R=0$ background,
\begin{equation}
 \delta E^{\rm full}[h]=\delta E^{\rm EW}[h]
  +2\beta(\bR_{\mu\nu}\phi+\bg_{\mu\nu}\bBox\phi
                                      -\bn_\mu\bn_\nu\phi).
\end{equation}
Four-dimensional conformal covariance of the Bach tensor gives
$\delta B^{\rm Bach}[\sigma\bg]=-\sigma\bar B^{\rm Bach}$.
Combining it with the on-shell equation gives
\begin{equation}
 \delta E^{\rm EW}[\sigma\bg]
  =\bg_{\mu\nu}\bBox\sigma-\bn_\mu\bn_\nu\sigma
                                             +\sigma\bR_{\mu\nu}.
\end{equation}
Therefore \eqref{eq:hhat} follows with
$\delta R[\widehat h]=\phi-6\beta\bBox\phi=0$. In invariant
amplitudes
\begin{equation}
 \widehat U=U-2\beta\Phi,\quad
 \widehat V=V+2\beta\Phi,\quad
 \widehat W=W+2\beta\Phi.
\end{equation}
The conformal contribution to the numerator of \eqref{eq:KI} is zero
identically. This is an exact linear scalar separation on the on-shell
background, not the perturbative metric redefinition of
Appendix~\ref{app:frames}.

The static radial trace equation is
\begin{equation}
 \left(r^2\sqrt{AB}\,\Phi'\right)'
 -r^2\sqrt{A/B}\left(m_0^2+\frac6{r^2}\right)\Phi=0.
\end{equation}
Multiplying by $\Phi^*$, integrating and taking the real part yields
\begin{align}\label{eq:scalarenergy}
 \left[\operatorname{Re}(\Phi^*r^2\sqrt{AB}\Phi')\right]_{r_h}^{\infty}
 =\int_{r_h}^{\infty}\!\dd r\,\bigg[
 r^2\sqrt{AB}|\Phi'|^2
 +r^2\sqrt{A/B}\left(m_0^2+\frac6{r^2}\right)|\Phi|^2\bigg].
\end{align}
An analytic nonextremal horizon gives $\sqrt{AB}=O(r-r_h)$ with
finite $\Phi,\Phi'$, so the lower boundary term vanishes. No applied
scalar tide selects the decaying massive branch at the upper endpoint,
where the term also vanishes. With $A,B>0$ outside and $m_0^2>0$,
the integral is nonnegative and forces $\Phi=0$. Horizon regularity
alone supplies only the lower-endpoint statement. The same static
integral with $\ell=0$ applied to the background trace gives the
usual $R=0$ result under these assumptions \cite{Lu:2015cqa,Lu:2015psa}.

For comparison, setting $U_0=r\Phi$ and
$\dd r_*/\dd r=(AB)^{-1/2}$ gives
\begin{equation}
 \ddot U_0-V_0U_0=0,\qquad
 V_0=A\left(m_0^2+\frac6{r^2}\right)+\frac{(AB)'}{2r}.
\end{equation}
The integral proof is preferable to assuming that this potential is
pointwise positive on every background.

\section{Details of the gauge-invariant field equations}\label{app:operator}
\subsection{Gauge changes and recovery of the metric}
For the stationary gauge covector
$\xi_t=TY$, $\xi_r=F_rY$, $\xi_A=r^2LD_AY$, the convention
$h\mapsto h-2\bn_{(\mu}\xi_{\nu)}$ gives
\begin{align}
 \Delta H_0&=BuF_r,&\Delta H_1&=-T'+uT,&
 \Delta H_2&=-2BF_r'-B'F_r,\\
 \Delta j_0&=-T,&\Delta J&=-F_r-r^2L',&
 \Delta K&=6L-2BF_r/r,&\Delta G_{\rm met}&=-L.
\end{align}
Thus $\Delta p_r=-F_r$, which verifies each cancellation in
\eqref{eq:invariants}. With $p_t=j_0$ and
$\mathsf P_a=p_aY$, $\mathsf P^A=G_{\rm met}D^AY$,
\begin{equation}
 H^{\GI}=h-\Lie_{\mathsf P}\bg.
\end{equation}
Since $\bar E=0$,
$\delta E[H^{\GI}]=\delta E[h]-\Lie_{\mathsf P}\bar E=\delta E[h]$.
The invariant representative is not an additional metric gauge choice.
After solving for $(U,V,W)$, arbitrary gauge functions reconstruct
\begin{align}\label{eq:reconstruct}
 H_0&=U-Bup_r,&H_2&=V+2Bp_r'+B'p_r,\nonumber\\
 K&=W-6G_{\rm met}+2Bp_r/r,&J&=p_r+r^2G_{\rm met}',\nonumber\\
 H_1&=j_0'-uj_0.
\end{align}
This is the even-time block $S=0$. The historical gauge-invariant
Schwarzschild construction provides the Ricci-flat positive control
\cite{Regge:1957td,Zerilli:1970se,Moncrief:1974am,Kodama:2003jz}.

\subsection{Linearizing the metric field equation}
For a metric representative $H_{\mu\nu}$ define
\begin{align}
 C^\lambda{}_{\mu\nu}[H]&=\tfrac12\bg^{\lambda\sigma}
 (\bn_\mu H_{\nu\sigma}+\bn_\nu H_{\mu\sigma}-\bn_\sigma H_{\mu\nu}),\\
 \Ric_{\mu\nu}[H]&=\bn_\lambda C^\lambda{}_{\mu\nu}
                                      -\bn_\nu C^\lambda{}_{\mu\lambda},\\
 \varphi[H]&=\bg^{\mu\nu}\Ric_{\mu\nu}[H]-H^{\mu\nu}\bR_{\mu\nu}.
\end{align}
The required curvature variations are
\begin{align}
 \delta R_{\mu\rho\nu\sigma}
 &=H_{\mu\lambda}\bR^\lambda{}_{\rho\nu\sigma}
 +\bg_{\mu\lambda}(\bn_\nu C^\lambda{}_{\sigma\rho}
                                      -\bn_\sigma C^\lambda{}_{\nu\rho}),\\
 \delta R^{\rho\sigma}
 &=\Ric^{\rho\sigma}[H]-H^{\rho\lambda}\bR_\lambda{}^\sigma
                                -H^{\sigma\lambda}\bR^\rho{}_\lambda,\\
 \delta\mathscr S
 &=2\bR^{\rho\sigma}\Ric_{\rho\sigma}[H]
                  -2H^{\rho\sigma}\bR_{\rho\lambda}\bR_\sigma{}^\lambda.
\end{align}
Write $\delta(\Box R_{\mu\nu})=\bBox\Ric_{\mu\nu}+\mathscr B_{\mu\nu}$,
where
\begin{align}
 \mathscr B_{\mu\nu}={}&-H^{\rho\sigma}\bn_\rho\bn_\sigma\bR_{\mu\nu}
 -\bg^{\rho\sigma}C^\lambda{}_{\rho\sigma}\bn_\lambda\bR_{\mu\nu}\nonumber\\
 &-2\bg^{\rho\sigma}C^\lambda{}_{\rho\mu}\bn_\sigma\bR_{\lambda\nu}
  -2\bg^{\rho\sigma}C^\lambda{}_{\rho\nu}\bn_\sigma\bR_{\mu\lambda}\nonumber\\
 &-\bg^{\rho\sigma}(\bn_\rho C^\lambda{}_{\sigma\mu})\bR_{\lambda\nu}
  -\bg^{\rho\sigma}(\bn_\rho C^\lambda{}_{\sigma\nu})\bR_{\mu\lambda}.
\end{align}
Linearizing \eqref{eq:full} on $\bR=0$ gives
\begin{align}\label{eq:metriccov}
 0={}&\Ric_{\mu\nu}-\tfrac12\bg_{\mu\nu}\varphi
 -2\alpha(\bBox\Ric_{\mu\nu}+\mathscr B_{\mu\nu})\nonumber\\
 &-4\alpha[(\delta R_{\mu\rho\nu\sigma})\bR^{\rho\sigma}
                              +\bR_{\mu\rho\nu\sigma}\delta R^{\rho\sigma}]
 +\alpha H_{\mu\nu}\bar{\mathscr S}+\alpha\bg_{\mu\nu}\delta\mathscr S
 \nonumber\\
 &+\tfrac23(\alpha-3\beta)\bn_\mu\bn_\nu\varphi
 +\tfrac13(\alpha+6\beta)\bg_{\mu\nu}\bBox\varphi
 +2(\beta+2\alpha/3)\bR_{\mu\nu}\varphi.
\end{align}
All curvature perturbations here are metric derivatives. Hence this is
the direct fourth-order formulation, without an unconstrained test
massive tensor.

\subsection{Component equations and constraints}
Normalize a polar tensor by
$T_{tt}=AT_0Y$, $T_{rr}=T_2Y/B$, $T_{rA}=T_rD_AY$,
$T_{AB}=r^2(T_K\gamma_{AB}Y+2T_GY_{AB})$.
For $\bm z=(U,V,W)^T$ and
$\bm Q=(Q_0,Q_2,\eta,Q_K,Q_G)^T$, define
\begin{align}\label{eq:arrays}
 \bm Q&=\sum_{k=0}^2R_k\bm z^{(k)},&
 \Phi&=\sum_{k=0}^2N_k\bm z^{(k)},\nonumber\\
 \bm{\delta E}&=\sum_{j=0}^2
       (H_j\bm z^{(j)}+E_j\bm Q^{(j)}+S_j\Phi^{(j)}).
\end{align}
The complete arrays are given in Appendix~\ref{app:metric}.
They determine a direct metric operator
\begin{equation}\label{eq:Pformula}
 \bm{\delta E}=\sum_{n=0}^4P_n\bm z^{(n)},\qquad
 P_n=H_n+\sum_{j,k=0}^2\sum_{s=0}^j
 \delta_{n,k+s}\binom js(E_jR_k^{(j-s)}+S_jN_k^{(j-s)}),
\end{equation}
with $H_n=0$ for $n>2$. There are five component rows with differential
identities, not five independent fourth-order evolutions.

In the auxiliary formulation the defining equation, scalar definition
and linearized Bianchi identity are retained:
\begin{align}\label{eq:constraints}
 q_{\mu\nu}&=\delta\operatorname{Ric}_{\mu\nu}[h],\\
 \phi&=\bg^{\mu\nu}q_{\mu\nu}-h^{\mu\nu}\bR_{\mu\nu},\\
 \tfrac12\partial_\nu\phi&=\bn^\mu q_{\mu\nu}
 -h^{\mu\rho}\bn_\rho\bR_{\mu\nu}
 -\bg^{\mu\rho}C^\lambda{}_{\rho\mu}\bR_{\lambda\nu}
 -\bg^{\mu\rho}C^\lambda{}_{\rho\nu}\bR_{\mu\lambda}.
\end{align}
Inserting their invariant representatives yields \eqref{eq:state}.
To reconstruct all Ricci components from the six-state solution, set
\begin{equation}
 Z=\begin{pmatrix}1&0&0&0&0&0\\
 1&0&0&0&-4r^2&0\\0&1&0&0&0&0\end{pmatrix},\qquad\bm z=Z\bm s.
\end{equation}
Then
\begin{equation}\label{eq:allRicciReconstruction}
 \bm Q=\left[R_0Z+R_1(Z'+ZL)
 +R_2\{Z''+2Z'L+Z(L'+L^2)\}\right]\bm s.
\end{equation}
This fixes $Q_0,Q_K$ as well as the retained components and enforces
$-Q_0+P_2+2Q_K+\rho_tU-\rho_rV-2\rho_oW=0$.

\subsection{The radial reduction and the order of a single KI equation}
Write $\Psi=c_0\bm s$. Its nonzero row entries are
\begin{equation}
 c_0=\frac1{12f_{\rm ref}'H}
 (-(Q_X+Q_Y),\ Q_X-Q_Y,\ 0,\ 0,\ 4r^2Q_X,\ 0).
\end{equation}
Let $T$ have rows $c_0,e_4^T,e_5^T,e_2^T,e_3^T,e_6^T$, where
$e_i$ denotes a one-based unit vector. Partition
\begin{equation}
 \begin{pmatrix}\bm y\\\bm w\end{pmatrix}=T\bm s,\quad
 \bm w=(W,P_2,\Gamma')^T,\quad
 (T'+TL)T^{-1}=\begin{pmatrix}C&\mathcal M\\N&F\end{pmatrix}.
\end{equation}
On a patch with $\det\mathcal M\ne0$ define
\begin{equation}
 J=C\mathcal M+\mathcal M'+\mathcal M F,\quad
 K=C'+C^2+\mathcal M N,\quad
 P=-J\mathcal M^{-1},\quad Q=J\mathcal M^{-1}C-K.
\end{equation}
This derives \eqref{eq:three}. If $c=\sqrt{AB}$, then
$P_*=cP-c'\Id$, $Q_*=c^2Q$. For
$\bm y=S_{\rm tr}\bm\Upsilon$ and
$\dot S_{\rm tr}=-P_*S_{\rm tr}/2$,
\begin{equation}
 \mathcal V=S_{\rm tr}^{-1}
 \left(\tfrac12\dot P_*+\tfrac14P_*^2-Q_*\right)S_{\rm tr}.
\end{equation}
Both the derivative-removing matrix and its similarity action matter.
At a zero of a selected velocity-map determinant use another complement
or the common six-state system. For the baseline, one such zero occurs
near $r/r_h=1.93491617$. The apparent $ru-2$ denominator occurs near
$1.43692441$ and is removed with the unsolved on-shell background
equation. Neither is a physical singularity.

An autonomous equation for the original $\Psi$ is a different question.
Define $c_{j+1}=c_j'+c_jL$, and let $\mathcal O_6$ have rows
$c_0,\ldots,c_5$. Where $\det\mathcal O_6\ne0$,
\begin{equation}
 \Psi^{(6)}-\sum_{j=0}^5a_j\Psi^{(j)}=0,
 \qquad(a_0,\ldots,a_5)=c_6\mathcal O_6^{-1}.
\end{equation}
The exact local order test uses
\begin{equation}\label{eq:localdatum}
 r=\tfrac72,\ A=\tfrac45,\ B=\tfrac34,\
 A'=\tfrac1{10},\ B'=\tfrac1{20},\ \mu=\tfrac65,\ M_{\rm ref}=1
\end{equation}
and finds
\begin{equation}
 \det\mathcal O_6=
 \frac{33665945368535977842506993772382729}
 {162502848000000000000000000}\ne0.
\end{equation}
Thus no universal closed equation of order five or less for this
observable can hold on the unrestricted generic local solution space.
This exact local datum is not a globally shot black hole and its
reference mass is not an ADM determination. The test does not prove
that the rank is six everywhere on every selected global branch.

\section{Checking the Schwarzschild tidal solution}\label{app:Schwarzschild}
On Ricci-flat Schwarzschild, the massive curvature equation and
$\delta\operatorname{Ric}[q]=\Delta_Lq/2$ prove
\eqref{eq:Schsplit}. Applying the KI observable to $h^{(0)}$ gives
\begin{equation}\label{eq:Zerilli}
 \ddot\Psi^{(0)}-V_Z\Psi^{(0)}=0,\qquad
 V_Z=\frac{2f(12r^3+12Mr^2+18M^2r+9M^3)}{r^3(2r+3M)^2}.
\end{equation}
Equivalently, the massless $U=V=H_2^{\rm rad}$ obeys
\begin{equation}
 (H_2^{\rm rad})''+\frac{2(r-M)}{r(r-2M)}(H_2^{\rm rad})'
 -\left[\frac6{r(r-2M)}+\frac{4M^2}{r^2(r-2M)^2}\right]
                                      H_2^{\rm rad}=0.
\end{equation}
With $z=r/M-1$ its independent solutions are proportional to
$(z^2-1)P_2''(z)$ and $(z^2-1)Q_2''(z)$.
The former gives \eqref{eq:Schpoly}; the latter is the response branch
and is singular at $z=1$. These statements are the static quadrupolar
specialization of the standard Ricci-flat construction
\cite{Binnington:2009bb,Damour:2009vw,Kodama:2003jz}.

The invariant representative in the ingoing coordinate $v=t+r_*$ has
\begin{equation}\label{eq:EFmetric}
 h_{vv}/Y=AU,\qquad h_{vr}/Y=-\sqrt{A/B}U,\qquad
 h_{rr}^{\rm EF}/Y=(U+V)/B.
\end{equation}
For the Schwarzschild polynomial, these become
$(r-2M)^2$, $-r(r-2M)$ and $2r^2$, all finite at the horizon.
$W=r^2-2M^2$ is also finite. Its $r^{-3}$ massless response is
exactly zero. Decaying massive fields behave exponentially times
radial powers and do not change this massless coefficient. The
argument assumes the specified no-growing-massive-source boundary
condition; it does not require asserting uniqueness of every
massive homogeneous boundary problem.

\section{Series solutions used in the tidal matching}\label{app:matching}
\subsection{Three regular solutions at the horizon}
For the baseline $r_h=1$, set $x=r-1$ and expand the regularized
six-state operator as
\begin{equation}
 L=\frac{R_{\rm hor}}x+\sum_{k\ge0}L_kx^k,
 \qquad\det(\rho\Id-R_{\rm hor})=\rho^3(\rho+1)^3.
\end{equation}
The three-dimensional kernel supplies the regular leading data
$(W_h,P_{2h},\Gamma_h)$. For the matrix of analytic horizon columns
$H(x)=\sum_{n\ge0}H_nx^n$,
\begin{equation}\label{eq:horrec}
 (n\Id-R_{\rm hor})H_n=
       \sum_{k=0}^{n-1}L_kH_{n-1-k},\qquad n\ge1.
\end{equation}
The determinant on the left is $n^3(n+1)^3$, so the recursion uniquely
extends each regular kernel vector. The complementary singular
solutions are excluded. At $r_h=1$, regular columns satisfy
$U_h=2\Gamma_h$, $V_h=-U_h$ and $Q_{0h}+P_{2h}=0$.
The cancellations are checked using the metric and curvature tensors
in ingoing coordinates, not only a master-field amplitude.

\subsection{The source and response solutions at large radius}
Use the asymptotic physical mass in $f=1-2M/r$; this is distinct from
the fixed auxiliary $M_{\rm ref}$. The normalized massless source is
\eqref{eq:Schpoly}. The response is
\begin{align}\label{eq:responsecol}
 Q_2(z)&=\frac{3z^2-1}{4}\log\frac{z+1}{z-1}-\frac32z,
 &z&=r/M-1,\\
 U_R=V_R&=\frac{5}{8M^3}(z^2-1)Q_2''(z)
                        =r^{-3}+O(r^{-4}),\\
 W_R&=\frac{fr^2u}{4}
       \left[U_R'+\frac{fr^2u^2+2fru+4}{fr^2u}U_R\right],
 &u&=f'/f.
\end{align}
These exact source and response solutions fix leading physical metric
powers; propagation away from the endpoint uses the full hairy operator.

For the massive curvature amplitudes $\bm w=(Q_0,\Gamma)^T$ the
asymptotic Schwarzschild reduction gives
$\bm w''+P_m\bm w'+Q_m\bm w=0$. In $m_2=1$ units,
\begin{equation}
 P_m=2\Id/r+O(r^{-2}),\qquad
 Q_m=-\Id-2M\Id/r+O(r^{-2}).
\end{equation}
The two-column expansions for each sign are
\begin{equation}
 \bm w_\sigma=e^{\sigma r}r^{p_\sigma}
            \sum_{n\ge0}V_n^\sigma r^{-n},\qquad
 p_\sigma=-1+\sigma M,\quad V_0^\sigma=\Id_2,\quad\sigma=\pm1.
\end{equation}
Writing $F=\sum V_nr^{-n}$, the residual after removing the prefactor is
\begin{align}
 \mathfrak R_\sigma[F]={}&F''+
     [2(\sigma+p_\sigma/r)\Id+P_m]F'\nonumber\\
 &+[(\sigma+p_\sigma/r)^2\Id-p_\sigma\Id/r^2
                   +(\sigma+p_\sigma/r)P_m+Q_m]F.
\end{align}
At order $r^{-n-1}$ the unknown $V_n$ has coefficient
$-2\sigma n\Id$. If $B_n$ is the known residual from lower orders,
then $V_n=B_n/(2\sigma n)$. Restoring units gives the prefactor
$e^{\sigma m_2r}(m_2r)^{-1+\sigma m_2M}$.

One obtains $P_m,Q_m$ directly from the explicit $L$: specialize to
Schwarzschild and restrict to its closed curvature block on
$(P_2,\eta,\Gamma,\Gamma')$. Transform this four-state system to
$(Q_0,Q_0',\Gamma,\Gamma')$ using
\eqref{eq:allRicciReconstruction}, then select its two acceleration
rows. This prescription specifies the endpoint matrices without
introducing an independent test-field approximation in the interior.
The massive metric columns at the endpoint follow from
$h^{(m)}=-2q/m_2^2$ and the same invariant map.
The neglected exponentially small background-hair corrections are a
finite-endpoint approximation. Increasing the endpoint is necessary
even when many inverse-radius terms and high precision are used.

\subsection{Matching and numerical accuracy}
Let $H(r)$ be the three regular horizon columns and propagate
$(T,R,D_1,D_2)$ inward. At $r_m$, solve
\begin{equation}\label{eq:matching}
 [H,-R,-D_1,-D_2]
 \begin{pmatrix}\bm c\\b\\d_1\\d_2\end{pmatrix}=T.
\end{equation}
This is a six-by-six linear problem with $a=1$. A nonzero homogeneous
solution would make the matching matrix singular and requires a
separate zero-mode analysis. Monitoring its conditioning is therefore
relevant near a candidate response pole.

The reference calculation used $R=40$, $r_m=6$, horizon offset $0.0004$,
order-16 perturbation horizon series, order-22 high-precision Taylor
propagation and 28-term massive asymptotic columns. The recovered
horizon data are approximately
\begin{equation}
 W_h=0.6176932032,\quad P_{2h}=0.4612016446,\quad
 \Gamma_h=0.01546179229.
\end{equation}
The outer decaying amplitudes depend on endpoint column normalization;
small endpoint-scaled coefficients are not directly physical Yukawa
charges. Moving $r_m$ to 8, changing the startup and expanding the
endpoint provided separate checks. The physical reconstructed state
was substituted into all thirteen original projected equations and
constraints and the five direct fourth-order metric rows. In the
recorded original reference verification, normalized original-equation
residuals were of order $10^{-32}$ or smaller; the independently
prepared neighboring runs have their own records. These are residual
diagnostics, not response error estimates.

\section{Numerical checks and limits of the scan}\label{app:provenance}
The accuracy tests distinguish three sources of numerical error. The
horizon series and its starting offset control the inner boundary.
The asymptotic series and outer endpoint control the far-field
approximation. Propagation and matching introduce integration and
linear-algebra errors. A small residual in one part of the calculation
does not bound all three.

Tables~\ref{tab:cutoff} and \ref{tab:nearby} give the documented
outer-endpoint tests for the reference point and two neighboring
couplings. The full scan is shown without per-point error bars.
We therefore use it to describe the branch trend, but do not fit a
low-mass divergence or locate a response pole from the curve. Such
claims require additional endpoint tests and matching diagnostics
at the relevant couplings.

The background comparison uses the numerical samples of the
normalized lapse and radial metric function through $r/r_h=40$.
No analytic fit is used in that figure. The reference response uses
the same physical source/response convention as the neighboring
runs. The original projected equations, Ricci definitions and Bianchi
constraints provide checks outside the reduced propagation system.

The scan holds $r_h$ fixed and varies $\widehat\alpha$. Its endpoints
are not bounds on all black holes of the theory. For $\alpha<0$,
$m_2^2<0$, and the massive static fields become oscillatory instead
of exponential. The positive-$\alpha$ tidal boundary prescription
then needs to be replaced; no negative-$\alpha$ hairy datum is
included in this scan.

\section{Component coefficients of the fourth-order metric equations}
\label{app:metric}
Use the row order $(tt,rr,rA,\hbox{angular trace},\hbox{angular TF})$ and
the conventions of \eqref{eq:arrays}. Matrices in this appendix use zero-based
entry labels $(i,j)$ to match their displayed tables. Columns of $R_k,H_k$
and $N_k$ correspond to $(U,V,W)$; columns of $E_k$ correspond to
$(Q_0,Q_2,\eta,Q_K,Q_G)$. All unlisted entries are zero.
The shorthand $B,u,v,\rho_t,\rho_r,\rho_o$ and primes mean the background
functions of \eqref{eq:rho}. The quantities $u',v'$ are derivatives, not
independent coefficients. The matrices are evaluated on $\bR=0$ and the
background equations. Their derivative compositions in \eqref{eq:Pformula}
retain all background-curvature couplings.

\subsection{\texorpdfstring{Matrices $R_0,R_1,R_2$}{Matrices R0, R1, R2}}

\begin{longtable}{@{}p{25mm}p{128mm}@{}}
\toprule Entry & Explicit value \\ \midrule\endfirsthead
\toprule Entry & Explicit value (continued) \\ \midrule\endhead
\bottomrule\endfoot
$(R_0)_{0,0}$ & $\displaystyle - \frac{B r^{2} u^{2} + B r^{2} u v + 2 B r^{2} u^{\prime} + 4 B r u - 12}{4 r^{2}}$ \\[6pt]
$(R_0)_{0,1}$ & $\displaystyle - \frac{B \left(r u^{2} + r u v + 2 r u^{\prime} + 4 u\right)}{4 r}$ \\[6pt]
$(R_0)_{1,1}$ & $\displaystyle \frac{3}{r^{2}}$ \\[6pt]
$(R_0)_{2,0}$ & $\displaystyle - \frac{- r u + 2}{4 r}$ \\[6pt]
$(R_0)_{2,1}$ & $\displaystyle \frac{r u + 2}{4 r}$ \\[6pt]
$(R_0)_{3,0}$ & $\displaystyle - \frac{3}{2 r^{2}}$ \\[6pt]
$(R_0)_{3,1}$ & $\displaystyle \frac{B r u + B r v + 2 B + 3}{2 r^{2}}$ \\[6pt]
$(R_0)_{3,2}$ & $\displaystyle - \frac{B r u + B r v + 2 B - 6}{2 r^{2}}$ \\[6pt]
$(R_0)_{4,0}$ & $\displaystyle \frac{1}{4 r^{2}}$ \\[6pt]
$(R_0)_{4,1}$ & $\displaystyle - \frac{1}{4 r^{2}}$ \\[6pt]
$(R_1)_{0,0}$ & $\displaystyle - \frac{B \left(2 r u + r v + 4\right)}{4 r}$ \\[6pt]
$(R_1)_{0,1}$ & $\displaystyle - \frac{B u}{4}$ \\[6pt]
$(R_1)_{0,2}$ & $\displaystyle \frac{B u}{2}$ \\[6pt]
$(R_1)_{1,0}$ & $\displaystyle \frac{B \left(2 u + v\right)}{4}$ \\[6pt]
$(R_1)_{1,1}$ & $\displaystyle \frac{B \left(r u + 4\right)}{4 r}$ \\[6pt]
$(R_1)_{1,2}$ & $\displaystyle - \frac{B \left(r v + 4\right)}{2 r}$ \\[6pt]
$(R_1)_{2,0}$ & $\displaystyle \frac{1}{2}$ \\[6pt]
$(R_1)_{2,2}$ & $\displaystyle - \frac{1}{2}$ \\[6pt]
$(R_1)_{3,0}$ & $\displaystyle \frac{B}{2 r}$ \\[6pt]
$(R_1)_{3,1}$ & $\displaystyle \frac{B}{2 r}$ \\[6pt]
$(R_1)_{3,2}$ & $\displaystyle - \frac{B \left(r u + r v + 8\right)}{4 r}$ \\[6pt]
$(R_2)_{0,0}$ & $\displaystyle - \frac{B}{2}$ \\[6pt]
$(R_2)_{1,0}$ & $\displaystyle \frac{B}{2}$ \\[6pt]
$(R_2)_{1,2}$ & $\displaystyle - B$ \\[6pt]
$(R_2)_{3,2}$ & $\displaystyle - \frac{B}{2}$ \\[6pt]
\end{longtable}

\subsection{\texorpdfstring{Matrices $N_0,N_1,N_2$}{Matrices N0, N1, N2}}

\begin{longtable}{@{}p{25mm}p{128mm}@{}}
\toprule Entry & Explicit value \\ \midrule\endfirsthead
\toprule Entry & Explicit value (continued) \\ \midrule\endhead
\bottomrule\endfoot
$(N_0)_{0,0}$ & $\displaystyle \rho_{t} + \frac{B r^{2} u^{2} + B r^{2} u v + 2 B r^{2} u^{\prime} + 4 B r u - 12}{4 r^{2}} - \frac{3}{r^{2}}$ \\[6pt]
$(N_0)_{0,1}$ & $\displaystyle \begin{aligned} &\frac{B u^{2}}{4} +\frac{B u v}{4} +\frac{B u^{\prime}}{2} -\rho_{r} +\frac{2 B u}{r} +\frac{B v}{r}\\[-1pt] &+\frac{2 B}{r^{2}} +\frac{6}{r^{2}}\end{aligned}$ \\[6pt]
$(N_0)_{0,2}$ & $\displaystyle - 2 \rho_{o} - \frac{B r u + B r v + 2 B - 6}{r^{2}}$ \\[6pt]
$(N_1)_{0,0}$ & $\displaystyle B \left(\frac{u}{2} + \frac{v}{4} + \frac{2 r u + r v + 4}{4 r} + \frac{1}{r}\right)$ \\[6pt]
$(N_1)_{0,1}$ & $\displaystyle B \left(\frac{u}{4} + \frac{r u + 4}{4 r} + \frac{1}{r}\right)$ \\[6pt]
$(N_1)_{0,2}$ & $\displaystyle - \frac{B \left(u + \frac{r v + 4}{r} + \frac{r u + r v + 8}{r}\right)}{2}$ \\[6pt]
$(N_2)_{0,0}$ & $\displaystyle B$ \\[6pt]
$(N_2)_{0,2}$ & $\displaystyle - 2 B$ \\[6pt]
\end{longtable}

\subsection{\texorpdfstring{Matrices $H_0,H_1,H_2$}{Matrices H0, H1, H2}}

\begin{longtable}{@{}p{25mm}p{128mm}@{}}
\toprule Entry & Explicit value \\ \midrule\endfirsthead
\toprule Entry & Explicit value (continued) \\ \midrule\endhead
\bottomrule\endfoot
$(H_0)_{0,0}$ & $\displaystyle \frac{\begin{aligned} &B \rho_{r} \alpha r^{2} u v +2 B \rho_{r} \alpha r^{2} u^{\prime} +2 \rho_{o}^{2} \alpha r^{2}\\[-1pt] &+\rho_{r}^{2} \alpha r^{2} -\rho_{t}^{2} \alpha r^{2} +4 B \rho_{o} \alpha r u -12 \rho_{o} \alpha\\[-1pt] &-12 \rho_{t} \alpha\end{aligned}}{r^{2}}$ \\[6pt]
$(H_0)_{0,1}$ & $\displaystyle \frac{\begin{aligned} &B \rho_{t} \alpha r u^{2} +2 B \rho_{r} \alpha r u v -B \rho_{t}^{\prime} \alpha r u\\[-1pt] &+4 B \rho_{r} \alpha r u^{\prime} -B \rho_{t}^{\prime} \alpha r v -2 B \rho_{t}^{\prime\prime} \alpha r\\[-1pt] &+2 \rho_{r}^{2} \alpha r +4 B \rho_{o} \alpha u -4 B \rho_{t}^{\prime} \alpha\end{aligned}}{r}$ \\[6pt]
$(H_0)_{0,2}$ & $\displaystyle \frac{4 \rho_{o} \alpha \left(B u + \rho_{o} r\right)}{r}$ \\[6pt]
$(H_0)_{1,0}$ & $\displaystyle \rho_{t} \alpha \left(B u v + 2 B u^{\prime} + 2 \rho_{t}\right)$ \\[6pt]
$(H_0)_{1,1}$ & $\displaystyle \frac{\begin{aligned} &-B \rho_{r} \alpha r^{2} u^{2} +B \rho_{r}^{\prime} \alpha r^{2} u +B \rho_{r}^{\prime} \alpha r^{2} v\\[-1pt] &+2 B \rho_{r}^{\prime\prime} \alpha r^{2} +2 \rho_{o}^{2} \alpha r^{2} -\rho_{r}^{2} \alpha r^{2}\\[-1pt] &+\rho_{t}^{2} \alpha r^{2} +4 B \rho_{r}^{\prime} \alpha r -8 B \rho_{r} \alpha -12 \rho_{o} \alpha\\[-1pt] &-12 \rho_{r} \alpha\end{aligned}}{r^{2}}$ \\[6pt]
$(H_0)_{1,2}$ & $\displaystyle \frac{4 \rho_{o} \alpha \left(- B r v + 2 B - \rho_{o} r^{2}\right)}{r^{2}}$ \\[6pt]
$(H_0)_{2,0}$ & $\displaystyle \frac{\alpha \left(- \rho_{o} r u - 2 \rho_{o} - \rho_{r} r u + 2 \rho_{r} + 4 \rho_{t}\right)}{2 r}$ \\[6pt]
$(H_0)_{2,1}$ & $\displaystyle \frac{\begin{aligned} &-\rho_{o} \alpha r u +\rho_{r} \alpha r u -2 \rho_{t} \alpha r u -4 \rho_{o}^{\prime} \alpha r\\[-1pt] &+4 \rho_{r}^{\prime} \alpha r -6 \rho_{o} \alpha +10 \rho_{r} \alpha\end{aligned}}{2 r}$ \\[6pt]
$(H_0)_{2,2}$ & $\displaystyle - \frac{4 \rho_{o} \alpha}{r}$ \\[6pt]
$(H_0)_{3,0}$ & $\displaystyle \frac{2 \rho_{t} \alpha \left(B r u + \rho_{t} r^{2} + 3\right)}{r^{2}}$ \\[6pt]
$(H_0)_{3,1}$ & $\displaystyle \frac{\begin{aligned} &B \rho_{o}^{\prime} \alpha r^{2} u +B \rho_{o}^{\prime} \alpha r^{2} v\\[-1pt] &+2 B \rho_{o}^{\prime\prime} \alpha r^{2} -2 \rho_{r}^{2} \alpha r^{2} -2 B \rho_{t} \alpha r u\\[-1pt] &-4 B \rho_{r} \alpha r v +4 B \rho_{o}^{\prime} \alpha r -8 B \rho_{o} \alpha +8 B \rho_{r} \alpha\\[-1pt] &-6 \rho_{r} \alpha\end{aligned}}{r^{2}}$ \\[6pt]
$(H_0)_{3,2}$ & $\displaystyle \frac{\begin{aligned} &-2 \rho_{o}^{2} \alpha r^{2} +\rho_{r}^{2} \alpha r^{2} +\rho_{t}^{2} \alpha r^{2}\\[-1pt] &+2 B \rho_{t} \alpha r u +2 B \rho_{r} \alpha r v -4 B \rho_{r} \alpha -20 \rho_{o} \alpha\end{aligned}}{r^{2}}$ \\[6pt]
$(H_0)_{4,0}$ & $\displaystyle - \frac{\rho_{t} \alpha}{r^{2}}$ \\[6pt]
$(H_0)_{4,1}$ & $\displaystyle \frac{\rho_{r} \alpha}{r^{2}}$ \\[6pt]
$(H_1)_{0,0}$ & $\displaystyle \frac{\begin{aligned} &3 B \rho_{t} \alpha r u +B \rho_{r} \alpha r v +B \rho_{t} \alpha r v +3 B \rho_{t}^{\prime} \alpha r\\[-1pt] &+4 B \rho_{o} \alpha +4 B \rho_{t} \alpha\end{aligned}}{r}$ \\[6pt]
$(H_1)_{0,1}$ & $\displaystyle B \alpha \left(\rho_{r} u - \rho_{t}^{\prime}\right)$ \\[6pt]
$(H_1)_{0,2}$ & $\displaystyle 2 B \alpha \left(- \rho_{o} u + \rho_{t}^{\prime}\right)$ \\[6pt]
$(H_1)_{1,0}$ & $\displaystyle B \alpha \left(- 2 \rho_{r} u - \rho_{t} v + \rho_{r}^{\prime}\right)$ \\[6pt]
$(H_1)_{1,1}$ & $\displaystyle \frac{\begin{aligned} &B \rho_{r} \alpha r u -B \rho_{t} \alpha r u +B \rho_{r} \alpha r v +5 B \rho_{r}^{\prime} \alpha r\\[-1pt] &-4 B \rho_{o} \alpha +4 B \rho_{r} \alpha\end{aligned}}{r}$ \\[6pt]
$(H_1)_{1,2}$ & $\displaystyle \frac{2 B \alpha \left(\rho_{o} r v + 4 \rho_{r} - \rho_{r}^{\prime} r\right)}{r}$ \\[6pt]
$(H_1)_{2,0}$ & $\displaystyle - 2 \rho_{t} \alpha$ \\[6pt]
$(H_1)_{2,1}$ & $\displaystyle \alpha \left(- \rho_{o} + \rho_{r}\right)$ \\[6pt]
$(H_1)_{2,2}$ & $\displaystyle \alpha \left(3 \rho_{o} - \rho_{r}\right)$ \\[6pt]
$(H_1)_{3,0}$ & $\displaystyle \frac{B \alpha \left(- 2 \rho_{t} + \rho_{o}^{\prime} r\right)}{r}$ \\[6pt]
$(H_1)_{3,1}$ & $\displaystyle \frac{B \alpha \left(- 2 \rho_{r} + \rho_{o}^{\prime} r\right)}{r}$ \\[6pt]
$(H_1)_{3,2}$ & $\displaystyle \frac{\begin{aligned} &B \rho_{o} \alpha r u +B \rho_{t} \alpha r u +B \rho_{o} \alpha r v +B \rho_{r} \alpha r v\\[-1pt] &+2 B \rho_{o}^{\prime} \alpha r +12 B \rho_{o} \alpha\end{aligned}}{r}$ \\[6pt]
$(H_2)_{0,0}$ & $\displaystyle 2 B \alpha \left(\rho_{r} + \rho_{t}\right)$ \\[6pt]
$(H_2)_{1,0}$ & $\displaystyle - 2 B \rho_{t} \alpha$ \\[6pt]
$(H_2)_{1,1}$ & $\displaystyle 2 B \rho_{r} \alpha$ \\[6pt]
$(H_2)_{1,2}$ & $\displaystyle 4 B \rho_{o} \alpha$ \\[6pt]
$(H_2)_{3,2}$ & $\displaystyle 2 B \alpha \left(\rho_{o} + \rho_{r}\right)$ \\[6pt]
\end{longtable}

\subsection{\texorpdfstring{Matrices $E_0,E_1,E_2$}{Matrices E0, E1, E2}}

\begin{longtable}{@{}p{25mm}p{128mm}@{}}
\toprule Entry & Explicit value \\ \midrule\endfirsthead
\toprule Entry & Explicit value (continued) \\ \midrule\endhead
\bottomrule\endfoot
$(E_0)_{0,0}$ & $\displaystyle \frac{B \alpha r^{2} u^{2} + 2 \rho_{t} \alpha r^{2} + 12 \alpha + r^{2}}{r^{2}}$ \\[6pt]
$(E_0)_{0,1}$ & $\displaystyle - \alpha \left(B u v + 2 B u^{\prime} + 2 \rho_{r}\right)$ \\[6pt]
$(E_0)_{0,3}$ & $\displaystyle - \frac{4 \alpha \left(B u + \rho_{o} r\right)}{r}$ \\[6pt]
$(E_0)_{1,0}$ & $\displaystyle - \alpha \left(B u v + 2 B u^{\prime} + 2 \rho_{t}\right)$ \\[6pt]
$(E_0)_{1,1}$ & $\displaystyle \frac{B \alpha r^{2} u^{2} + 8 B \alpha + 2 \rho_{r} \alpha r^{2} + 12 \alpha + r^{2}}{r^{2}}$ \\[6pt]
$(E_0)_{1,2}$ & $\displaystyle - \frac{48 B \alpha}{r^{3}}$ \\[6pt]
$(E_0)_{1,3}$ & $\displaystyle \frac{4 \alpha \left(B r v - 2 B + \rho_{o} r^{2}\right)}{r^{2}}$ \\[6pt]
$(E_0)_{2,1}$ & $\displaystyle - \frac{4 \alpha}{r}$ \\[6pt]
$(E_0)_{2,2}$ & $\displaystyle \frac{\begin{aligned} &B \alpha r^{2} u^{2} -B \alpha r^{2} u v -2 B \alpha r^{2} v^{2} -2 B \alpha r^{2} v^{\prime}\\[-1pt] &+2 B \alpha r u -2 B \alpha r v +20 B \alpha +20 \alpha +2 r^{2}\end{aligned}}{2 r^{2}}$ \\[6pt]
$(E_0)_{2,3}$ & $\displaystyle \frac{4 \alpha}{r}$ \\[6pt]
$(E_0)_{2,4}$ & $\displaystyle - \frac{16 \alpha}{r}$ \\[6pt]
$(E_0)_{3,0}$ & $\displaystyle - \frac{2 \alpha \left(B u + \rho_{t} r\right)}{r}$ \\[6pt]
$(E_0)_{3,1}$ & $\displaystyle \frac{2 \alpha \left(B r v - 2 B + \rho_{r} r^{2}\right)}{r^{2}}$ \\[6pt]
$(E_0)_{3,2}$ & $\displaystyle \frac{24 B \alpha}{r^{3}}$ \\[6pt]
$(E_0)_{3,3}$ & $\displaystyle \frac{8 B \alpha + 4 \rho_{o} \alpha r^{2} + 8 \alpha + r^{2}}{r^{2}}$ \\[6pt]
$(E_0)_{4,2}$ & $\displaystyle - \frac{4 B \alpha}{r^{3}}$ \\[6pt]
$(E_0)_{4,4}$ & $\displaystyle \frac{8 \alpha + r^{2}}{r^{2}}$ \\[6pt]
$(E_1)_{0,0}$ & $\displaystyle - \frac{B \alpha \left(r u + r v + 4\right)}{r}$ \\[6pt]
$(E_1)_{1,1}$ & $\displaystyle - \frac{B \alpha \left(r u + r v + 4\right)}{r}$ \\[6pt]
$(E_1)_{2,2}$ & $\displaystyle - B \alpha \left(u + 3 v\right)$ \\[6pt]
$(E_1)_{3,3}$ & $\displaystyle - \frac{B \alpha \left(r u + r v + 4\right)}{r}$ \\[6pt]
$(E_1)_{4,4}$ & $\displaystyle - \frac{B \alpha \left(r u + r v + 4\right)}{r}$ \\[6pt]
$(E_2)_{0,0}$ & $\displaystyle - 2 B \alpha$ \\[6pt]
$(E_2)_{1,1}$ & $\displaystyle - 2 B \alpha$ \\[6pt]
$(E_2)_{2,2}$ & $\displaystyle - 2 B \alpha$ \\[6pt]
$(E_2)_{3,3}$ & $\displaystyle - 2 B \alpha$ \\[6pt]
$(E_2)_{4,4}$ & $\displaystyle - 2 B \alpha$ \\[6pt]
\end{longtable}

\subsection{\texorpdfstring{Matrices $S_0,S_1,S_2$}{Matrices S0, S1, S2}}

\begin{longtable}{@{}p{25mm}p{128mm}@{}}
\toprule Entry & Explicit value \\ \midrule\endfirsthead
\toprule Entry & Explicit value (continued) \\ \midrule\endhead
\bottomrule\endfoot
$(S_0)_{0,0}$ & $\displaystyle \frac{- 8 \rho_{t} \alpha r^{2} - 12 \rho_{t} \beta r^{2} + 12 \alpha + 72 \beta + 3 r^{2}}{6 r^{2}}$ \\[6pt]
$(S_0)_{1,0}$ & $\displaystyle \frac{8 \rho_{r} \alpha r^{2} + 12 \rho_{r} \beta r^{2} - 12 \alpha - 72 \beta - 3 r^{2}}{6 r^{2}}$ \\[6pt]
$(S_0)_{2,0}$ & $\displaystyle \frac{2 \left(- \alpha + 3 \beta\right)}{3 r}$ \\[6pt]
$(S_0)_{3,0}$ & $\displaystyle \frac{8 \rho_{o} \alpha r^{2} + 12 \rho_{o} \beta r^{2} - 24 \alpha - 36 \beta - 3 r^{2}}{6 r^{2}}$ \\[6pt]
$(S_0)_{4,0}$ & $\displaystyle \frac{\alpha - 3 \beta}{3 r^{2}}$ \\[6pt]
$(S_1)_{0,0}$ & $\displaystyle - \frac{B \left(3 \alpha r u + \alpha r v + 4 \alpha + 6 \beta r v + 24 \beta\right)}{6 r}$ \\[6pt]
$(S_1)_{1,0}$ & $\displaystyle \frac{B \left(\alpha r u + 3 \alpha r v + 4 \alpha + 6 \beta r u + 24 \beta\right)}{6 r}$ \\[6pt]
$(S_1)_{2,0}$ & $\displaystyle 2 \left(\frac{\alpha}{3} - \beta\right)$ \\[6pt]
$(S_1)_{3,0}$ & $\displaystyle \frac{B \left(\alpha r u + \alpha r v + 8 \alpha + 6 \beta r u + 6 \beta r v + 12 \beta\right)}{6 r}$ \\[6pt]
$(S_2)_{0,0}$ & $\displaystyle - B \left(\frac{\alpha}{3} + 2 \beta\right)$ \\[6pt]
$(S_2)_{1,0}$ & $\displaystyle B \alpha$ \\[6pt]
$(S_2)_{3,0}$ & $\displaystyle B \left(\frac{\alpha}{3} + 2 \beta\right)$ \\[6pt]
\end{longtable}

\section{Coefficients of the six-state gravitational system}\label{app:L}
The state is $(U,W,P_2,\eta,\Gamma,\Gamma')$ after scalar exclusion.
Here indices of $L$ are zero-based, $b=B$, $b'=B'$, $u=A'/A$, and
$\mu=m_2^2$. Every unlisted entry is zero. On a patch where a denominator
used in this elimination vanishes, use the unsolved component system.
All these variables are gauge invariants; no condition has been placed on
the original metric gauge coefficients.

\subsection{\texorpdfstring{Row 0 of $L$}{Row 0 of L}}

\begin{longtable}{@{}p{25mm}p{128mm}@{}}
\toprule Entry & Explicit value \\ \midrule\endfirsthead
\toprule Entry & Explicit value (continued) \\ \midrule\endhead
\bottomrule\endfoot
$L_{0,0}$ & $\displaystyle - \frac{B r^{2} u^{2} + 2 B r u + 4}{B r^{2} u}$ \\[6pt]
$L_{0,1}$ & $\displaystyle \frac{4}{B r^{2} u}$ \\[6pt]
$L_{0,2}$ & $\displaystyle \frac{2}{B u}$ \\[6pt]
$L_{0,3}$ & $\displaystyle \frac{2 \left(r u + 2\right)}{r u}$ \\[6pt]
$L_{0,4}$ & $\displaystyle \frac{2 \left(B r^{2} u^{2} + 4 B r u + 4 B - 4\right)}{B u}$ \\[6pt]
\end{longtable}

\subsection{\texorpdfstring{Row 1 of $L$}{Row 1 of L}}

\begin{longtable}{@{}p{25mm}p{128mm}@{}}
\toprule Entry & Explicit value \\ \midrule\endfirsthead
\toprule Entry & Explicit value (continued) \\ \midrule\endhead
\bottomrule\endfoot
$L_{1,0}$ & $\displaystyle - \frac{2 \left(B r u + 2\right)}{B r^{2} u}$ \\[6pt]
$L_{1,1}$ & $\displaystyle \frac{4}{B r^{2} u}$ \\[6pt]
$L_{1,2}$ & $\displaystyle \frac{2}{B u}$ \\[6pt]
$L_{1,3}$ & $\displaystyle \frac{4}{r u}$ \\[6pt]
$L_{1,4}$ & $\displaystyle \frac{4 \left(B r u + 2 B - 2\right)}{B u}$ \\[6pt]
\end{longtable}

\subsection{\texorpdfstring{Row 2 of $L$}{Row 2 of L}}

\begin{longtable}{@{}p{25mm}p{128mm}@{}}
\toprule Entry & Explicit value \\ \midrule\endfirsthead
\toprule Entry & Explicit value (continued) \\ \midrule\endhead
\bottomrule\endfoot
$L_{2,0}$ & $\displaystyle \frac{\begin{aligned} &-9 B^{3} r^{4} u^{4} +3 B^{3} r^{3} u^{3} +12 B^{3} r^{2} u^{2} +12 B^{2} B^{\prime} r^{3} u^{2}\\[-1pt] &+12 B^{2} B^{\prime} r^{2} u -6 B^{2} \mu r^{5} u^{3} -6 B^{2} \mu r^{4} u^{2} -16 B^{2} r^{3} u^{3}\\[-1pt] &-8 B^{2} r^{2} u^{2} +56 B^{2} r u +9 B \left(B^{\prime}\right)^{2} r^{4} u^{2}\\[-1pt] &+9 B \left(B^{\prime}\right)^{2} r^{3} u +8 B B^{\prime} r^{3} u^{2} +36 B B^{\prime} r^{2} u\\[-1pt] &+40 B B^{\prime} r +6 B \mu r^{4} u^{2} -16 B \mu r^{3} u -16 B \mu r^{2} +8 B r^{2} u^{2} -56 B r u +32 B\\[-1pt] &+36 \left(B^{\prime}\right)^{2} r^{2} -8 B^{\prime} r +16 \mu r^{2} -32\end{aligned}}{2 B r^{4} u \left(3 B r u + 3 B^{\prime} r + 4\right)}$ \\[6pt]
$L_{2,1}$ & $\displaystyle \frac{\begin{aligned} &-2 B^{2} r^{2} u^{2} -20 B^{2} r u -20 B B^{\prime} r^{2} u -20 B B^{\prime} r +8 B \mu r^{3} u\\[-1pt] &+8 B \mu r^{2} +20 B r u -16 B -18 \left(B^{\prime}\right)^{2} r^{2} +4 B^{\prime} r -8 \mu r^{2} +16\end{aligned}}{B r^{4} u \left(3 B r u + 3 B^{\prime} r + 4\right)}$ \\[6pt]
$L_{2,2}$ & $\displaystyle \frac{\begin{aligned} &-B^{2} r^{3} u^{3} -11 B^{2} r^{2} u^{2} -10 B^{2} r u -B B^{\prime} r^{3} u^{2} -20 B B^{\prime} r^{2} u\\[-1pt] &-10 B B^{\prime} r +B \mu r^{4} u^{2} +2 B \mu r^{3} u +4 B \mu r^{2} -6 B r u -8 B\\[-1pt] &-9 \left(B^{\prime}\right)^{2} r^{2} +2 B^{\prime} r -4 \mu r^{2} +8\end{aligned}}{B r^{2} u \left(3 B r u + 3 B^{\prime} r + 4\right)}$ \\[6pt]
$L_{2,3}$ & $\displaystyle \frac{\begin{aligned} &2 B^{2} r^{4} u^{4} +15 B^{2} r^{3} u^{3} -40 B^{2} r u +2 B B^{\prime} r^{4} u^{3}\\[-1pt] &+6 B B^{\prime} r^{3} u^{2} -36 B B^{\prime} r^{2} u -40 B B^{\prime} r -2 B \mu r^{5} u^{3}\\[-1pt] &+12 B \mu r^{4} u^{2} +12 B \mu r^{3} u +16 B \mu r^{2} +68 B r^{2} u^{2} +24 B r u -32 B\\[-1pt] &-9 \left(B^{\prime}\right)^{2} r^{3} u -36 \left(B^{\prime}\right)^{2} r^{2} +20 B^{\prime} r^{2} u\\[-1pt] &+8 B^{\prime} r -4 \mu r^{3} u -16 \mu r^{2} +32 r u +32\end{aligned}}{2 r^{3} u \left(3 B r u + 3 B^{\prime} r + 4\right)}$ \\[6pt]
$L_{2,4}$ & $\displaystyle \frac{\begin{aligned} &11 B^{3} r^{4} u^{4} -12 B^{3} r^{3} u^{3} -36 B^{3} r^{2} u^{2} -40 B^{3} r u\\[-1pt] &+2 B^{2} B^{\prime} r^{4} u^{3} -30 B^{2} B^{\prime} r^{3} u^{2} -72 B^{2} B^{\prime} r^{2} u\\[-1pt] &-40 B^{2} B^{\prime} r +4 B^{2} \mu r^{5} u^{3} +12 B^{2} \mu r^{4} u^{2} +24 B^{2} \mu r^{3} u\\[-1pt] &+16 B^{2} \mu r^{2} +18 B^{2} r^{3} u^{3} +20 B^{2} r^{2} u^{2} +48 B^{2} r u -32 B^{2}\\[-1pt] &-9 B \left(B^{\prime}\right)^{2} r^{4} u^{2} -18 B \left(B^{\prime}\right)^{2} r^{3} u\\[-1pt] &-36 B \left(B^{\prime}\right)^{2} r^{2} -14 B B^{\prime} r^{3} u^{2} +64 B B^{\prime} r^{2} u\\[-1pt] &+48 B B^{\prime} r -4 B \mu r^{4} u^{2} -24 B \mu r^{3} u -32 B \mu r^{2} -32 B r^{2} u^{2} +40 B r u\\[-1pt] &+64 B +36 \left(B^{\prime}\right)^{2} r^{2} -8 B^{\prime} r +16 \mu r^{2} -32\end{aligned}}{B r^{2} u \left(3 B r u + 3 B^{\prime} r + 4\right)}$ \\[6pt]
$L_{2,5}$ & $\displaystyle \frac{\begin{aligned} &-12 B^{2} r^{2} u^{2} -12 B^{2} r u -12 B B^{\prime} r^{2} u -12 B B^{\prime} r +4 B r^{2} u^{2} -20 B r u\\[-1pt] &+12 B^{\prime} r +16\end{aligned}}{3 B r u + 3 B^{\prime} r + 4}$ \\[6pt]
\end{longtable}

\subsection{\texorpdfstring{Row 3 of $L$}{Row 3 of L}}

\begin{longtable}{@{}p{25mm}p{128mm}@{}}
\toprule Entry & Explicit value \\ \midrule\endfirsthead
\toprule Entry & Explicit value (continued) \\ \midrule\endhead
\bottomrule\endfoot
$L_{3,0}$ & $\displaystyle \frac{\begin{aligned} &3 B^{3} r^{2} u^{3} +12 B^{3} r u^{2} +12 B^{2} B^{\prime} r^{2} u^{2} +12 B^{2} B^{\prime} r u\\[-1pt] &-6 B^{2} \mu r^{4} u^{3} -6 B^{2} \mu r^{3} u^{2} -4 B^{2} r^{2} u^{3} -8 B^{2} r u^{2} +32 B^{2} u\\[-1pt] &+9 B \left(B^{\prime}\right)^{2} r^{2} u -4 B B^{\prime} r^{2} u^{2} +12 B B^{\prime} r u +16 B B^{\prime}\\[-1pt] &+6 B \mu r^{3} u^{2} -16 B \mu r^{2} u -16 B \mu r +8 B r u^{2} -32 B u +12 \left(B^{\prime}\right)^{2} r\\[-1pt] &-16 B^{\prime} +16 \mu r\end{aligned}}{2 B^{2} r^{2} u \left(r u - 2\right) \left(3 B r u + 3 B^{\prime} r + 4\right)}$ \\[6pt]
$L_{3,1}$ & $\displaystyle \frac{\begin{aligned} &-2 B^{2} r u^{2} -8 B^{2} u -8 B B^{\prime} r u -8 B B^{\prime} +8 B \mu r^{2} u +8 B \mu r +8 B u\\[-1pt] &-6 \left(B^{\prime}\right)^{2} r +8 B^{\prime} -8 \mu r\end{aligned}}{B^{2} r^{2} u \left(r u - 2\right) \left(3 B r u + 3 B^{\prime} r + 4\right)}$ \\[6pt]
$L_{3,2}$ & $\displaystyle \frac{\begin{aligned} &2 B^{2} r^{2} u^{3} -5 B^{2} r u^{2} -4 B^{2} u +2 B B^{\prime} r^{2} u^{2} -8 B B^{\prime} r u\\[-1pt] &-4 B B^{\prime} +B \mu r^{3} u^{2} +2 B \mu r^{2} u +4 B \mu r +4 B r u^{2} -4 B u\\[-1pt] &-3 \left(B^{\prime}\right)^{2} r +4 B^{\prime} -4 \mu r\end{aligned}}{B^{2} u \left(r u - 2\right) \left(3 B r u + 3 B^{\prime} r + 4\right)}$ \\[6pt]
$L_{3,3}$ & $\displaystyle \frac{\begin{aligned} &-B^{2} r^{3} u^{4} -9 B^{2} r^{2} u^{3} +12 B^{2} r u^{2} -16 B^{2} u -4 B B^{\prime} r^{3} u^{3}\\[-1pt] &-6 B B^{\prime} r^{2} u^{2} -16 B B^{\prime} -2 B \mu r^{4} u^{3} +12 B \mu r^{3} u^{2} +12 B \mu r^{2} u\\[-1pt] &+16 B \mu r -4 B r^{2} u^{3} +12 B r u^{2} +16 B u -3 \left(B^{\prime}\right)^{2} r^{3} u^{2}\\[-1pt] &+3 \left(B^{\prime}\right)^{2} r^{2} u -12 \left(B^{\prime}\right)^{2} r -4 B^{\prime} r^{2} u^{2}\\[-1pt] &+12 B^{\prime} r u +16 B^{\prime} -4 \mu r^{2} u -16 \mu r\end{aligned}}{2 B r u \left(r u - 2\right) \left(3 B r u + 3 B^{\prime} r + 4\right)}$ \\[6pt]
$L_{3,4}$ & $\displaystyle \frac{\begin{aligned} &-B^{3} r^{3} u^{4} -6 B^{3} r^{2} u^{3} -12 B^{3} r u^{2} -16 B^{3} u -4 B^{2} B^{\prime} r^{3} u^{3}\\[-1pt] &-12 B^{2} B^{\prime} r^{2} u^{2} -24 B^{2} B^{\prime} r u -16 B^{2} B^{\prime} +4 B^{2} \mu r^{4} u^{3}\\[-1pt] &+12 B^{2} \mu r^{3} u^{2} +24 B^{2} \mu r^{2} u +16 B^{2} \mu r +20 B^{2} r^{2} u^{3} -20 B^{2} r u^{2}\\[-1pt] &+32 B^{2} u -3 B \left(B^{\prime}\right)^{2} r^{3} u^{2} -6 B \left(B^{\prime}\right)^{2} r^{2} u\\[-1pt] &-12 B \left(B^{\prime}\right)^{2} r +12 B B^{\prime} r^{2} u^{2} +8 B B^{\prime} r u +32 B B^{\prime}\\[-1pt] &-4 B \mu r^{3} u^{2} -24 B \mu r^{2} u -32 B \mu r -8 B r u^{2} +12 \left(B^{\prime}\right)^{2} r\\[-1pt] &-16 B^{\prime} +16 \mu r\end{aligned}}{B^{2} u \left(r u - 2\right) \left(3 B r u + 3 B^{\prime} r + 4\right)}$ \\[6pt]
$L_{3,5}$ & $\displaystyle \frac{4 r \left(r u - 2\right)}{3 B r u + 3 B^{\prime} r + 4}$ \\[6pt]
\end{longtable}

\subsection{\texorpdfstring{Row 4 of $L$}{Row 4 of L}}

\begin{longtable}{@{}p{25mm}p{128mm}@{}}
\toprule Entry & Explicit value \\ \midrule\endfirsthead
\toprule Entry & Explicit value (continued) \\ \midrule\endhead
\bottomrule\endfoot
$L_{4,5}$ & $\displaystyle 1$ \\[6pt]
\end{longtable}

\subsection{\texorpdfstring{Row 5 of $L$}{Row 5 of L}}

\begin{longtable}{@{}p{25mm}p{128mm}@{}}
\toprule Entry & Explicit value \\ \midrule\endfirsthead
\toprule Entry & Explicit value (continued) \\ \midrule\endhead
\bottomrule\endfoot
$L_{5,0}$ & $\displaystyle \frac{B u - B^{\prime}}{2 B r^{3}}$ \\[6pt]
$L_{5,3}$ & $\displaystyle - \frac{2}{r^{3}}$ \\[6pt]
$L_{5,4}$ & $\displaystyle \frac{- 2 B r u - 2 B + \mu r^{2} + 6}{B r^{2}}$ \\[6pt]
$L_{5,5}$ & $\displaystyle - \frac{B r u + 4 B + B^{\prime} r}{2 B r}$ \\[6pt]
\end{longtable}

\section{Attempts to separate the three gravitational equations}
\label{app:decoupling}
This appendix gives the transformation equations and the local tests
used to search for simpler gravitational variables. Its scope is the static three-component
system after scalar separation. The tests on generic local background
data do not assume that those data lie on a globally regular,
asymptotically flat black hole.

\subsection{Why diagonalizing the potential is not enough}
Let $x=r_*$ and $\bm\Upsilon=E(x)\bm z$, with
$E^{-1}\mathcal VE=\Lambda$ diagonal. Equation~\eqref{eq:matrixSch}
becomes
\begin{equation}
 \ddot{\bm z}+2E^{-1}\dot E\dot{\bm z}
       +(E^{-1}\ddot E-\Lambda)\bm z=0.
\end{equation}
The varying eigenbasis introduces a connection $E^{-1}\dot E$.
The off-diagonal terms generally survive. Removing them by a second
transport changes the potential again, so diagonalizing each potential
matrix at one radius is not an independent-channel reduction.

For a point map directly from a system with no first derivatives to
another such system, the coefficient of $\dot{\bm z}$ requires
$\dot E=0$. The potential family must therefore admit a constant
invariant splitting in that transported basis. At the local datum
\eqref{eq:localdatum}, the exact linear commutant test for
$\mathcal V$ and $\dot{\mathcal V}$ has rank eight in the nine
matrix entries. Only multiples of $\Id_3$ commute with both. This
rules out a nontrivial separation by a point transformation, including a one-plus-two block,
near that datum in the specified class. It says nothing by itself
about transformations involving derivatives.

\subsection{Allowing first derivatives in the new variables}
Seek
\begin{equation}\label{eq:dtrans}
 \bm w=F(x)\dot{\bm\Upsilon}+G(x)\bm\Upsilon,
 \qquad \ddot{\bm w}-\mathscr D(x)\bm w=0,
 \quad\mathscr D=\operatorname{diag}(d_1,d_2,d_3).
\end{equation}
Using $\ddot{\bm\Upsilon}=\mathcal V\bm\Upsilon$ twice gives the
necessary and sufficient conditions
\begin{align}\label{eq:FG}
 \ddot F+2\dot G+F\mathcal V-\mathscr DF&=0,\\
 \ddot G+2\dot F\mathcal V+F\dot{\mathcal V}
                         +G\mathcal V-\mathscr DG&=0.
\end{align}
Matrix ordering is essential. The correct invertibility condition is
that of the six-state map,
\begin{equation}\label{eq:phaseT}
 \begin{pmatrix}\bm w\\\dot{\bm w}\end{pmatrix}
 =\mathcal T\begin{pmatrix}\bm\Upsilon\\\dot{\bm\Upsilon}\end{pmatrix},
 \qquad\mathcal T=\begin{pmatrix}
 G&F\\\dot G+F\mathcal V&G+\dot F
 \end{pmatrix},\quad \det\mathcal T\ne0.
\end{equation}
Neither $\det F$ nor $\det G$ separately supplies this condition.

Higher finite derivative maps do not enlarge the on-solution linear
class: if $\bm\Upsilon^{(n)}=A_n\bm\Upsilon+B_n\dot{\bm\Upsilon}$,
then
\begin{equation}
 A_{n+1}=\dot A_n+B_n\mathcal V,\qquad
 B_{n+1}=A_n+\dot B_n,
 \qquad A_0=\Id,\quad B_0=0.
\end{equation}
This on-solution reduction is not an off-shell operator factorization.

\subsection{A formal local separation using integrated coefficients}
Define
\begin{equation}
 K_V=\begin{pmatrix}0&\Id\\\mathcal V&0\end{pmatrix},\qquad
 K_D=\begin{pmatrix}0&\Id\\\mathscr D&0\end{pmatrix}.
\end{equation}
Equation~\eqref{eq:FG} is equivalent to
$\dot{\mathcal T}=K_D\mathcal T-\mathcal TK_V$, with the block
relations \eqref{eq:phaseT}. Let $\Phi_V,\Phi_D$ be the respective
fundamental matrices normalized to $\Id_6$ at $x_0$. Then
\begin{equation}\label{eq:fundmap}
 \mathcal T=\Phi_D\Phi_V^{-1},\qquad \det\mathcal T=1.
\end{equation}
The determinant follows from the vanishing traces of $K_V,K_D$.
Thus, for any smooth chosen diagonal target, this map is invertible
on a regular interval. It is a mathematical existence result for
integrated coefficients, not a simple background-algebraic master-field
formula. Its boundary limits need not be regular or source preserving.

For free targets $\mathscr D=0$, set $x_0=0$. If
$\ddot X=\mathcal VX$, $X(0)=\Id$, $\dot X(0)=0$, and
$\ddot Y=\mathcal VY$, $Y(0)=0$, $\dot Y(0)=\Id$, then
\begin{equation}
 \Phi_0=\begin{pmatrix}\Id&x\Id\\0&\Id\end{pmatrix},\qquad
 \bm\Upsilon=X\bm w+(Y-xX)\dot{\bm w}.
\end{equation}
The free solution $\bm w=\bm a+x\bm b$ reproduces
$\bm\Upsilon=X\bm a+Y\bm b$. All original dynamics is encoded in
$X,Y$. Near $x=0$, the transformation begins as
\begin{equation}
 F=\frac{x^3}{3}\mathcal V_0+
          \frac{x^4}{12}\dot{\mathcal V}_0+O(x^5),\qquad
 G=\Id-\frac{x^2}{2}\mathcal V_0-
          \frac{x^3}{6}\dot{\mathcal V}_0+O(x^4).
\end{equation}
We also integrated these coefficient equations
on a local non-Ricci-flat background as a check. No identification of
the free target amplitudes with physical massless or massive tides
follows. Source/response extraction still requires the inverse map
and physical metric reconstruction.

\subsection{Tests of simpler transformations}
The strict ordinary intertwining condition for
$\mathscr U=F\partial_x+G$ is
\begin{equation}
 (\partial_x^2-\mathscr D)\mathscr U
                  =\mathscr U(\partial_x^2-\mathcal V).
\end{equation}
It requires $\dot F=0$. For a row targeting $\lambda_i$,
\begin{equation}
 \dot g=\tfrac12(\lambda_if-f\mathcal V),\qquad
 0=g(\mathcal V-\lambda_i\Id)
          +\tfrac12f(\dot{\mathcal V}+\dot\lambda_i\Id).
\end{equation}
The latter compatibility equation and its derivative give six linear
conditions on the six entries of $(f,g)$. At the test datum they have
full rank for the instantaneous-eigenvalue targets and the free
targets, leaving no invertible ordinary map to those specified
potentials. This does not exclude variable-$F$ maps such as
\eqref{eq:fundmap}, or ordinary maps to every other target. The scalar
and matrix Darboux literature gives the broader setting
\cite{Glampedakis:2017rar,Samsonov:2004ri}.

A genuine invariant solution-space projector in the six-state system
would satisfy
\begin{equation}
 J'=LJ-JL,\qquad J^2=J.
\end{equation}
The bounded ansatz in the finite search was $J=\sum_{k=1}^{20}f_kC_k$,
with constant six-by-six matrices $C_k$ and scalar span
\begin{align}
 \{f_k\}=\{&1,r,r^2,r^{-1},r^{-2},B,B^{-1},u,u^{-1},B',B'/B,
 \rho_t,\rho_r,\nonumber\\
 &r\rho_t,r\rho_r,r^2\rho_t,r^2\rho_r,
 (ru-2)^{-1},(3Bru+3rB'+4)^{-1},(r+3)^{-1}\}.
\end{align}
At thirty-six generic on-shell local data the linear parallel-map
conditions have rank 719 in 720 unknowns. The remaining solution is
a scalar multiple of $\Id_6$, giving no nontrivial projector in this
finite universal ansatz. The Schwarzschild massless projector is a
positive control within the same span when the background is
restricted to Schwarzschild. More complicated rational functions,
other targets and identities specific to a globally selected branch
remain outside this exclusion.

Finally, the directly continued candidate
$\Psi[h]+2\Psi[q^{\GI}]/m_2^2$ has observable rank two on the
Ricci-flat control but rank six at the generic non-Einstein test datum.
It is gauge invariant without being an autonomous second-order field.
These tests did not produce a simple geometric
massless separation has not been obtained on the L\"u branch.
Neither mathematical decoupling in a broad integrated class nor
universal impossibility of geometric separation is the appropriate
claim. The coupled system is sufficient for the tidal boundary problem.

\clearpage
\bibliographystyle{unsrtnat}
\bibliography{references}
\end{document}